\documentclass[aps,prd,
showkeys,showpacs,amssymb,cite,
amsfonts,epsf,preprintnumbers,nofootinbib,superscriptaddress
]{revtex4}

\usepackage[dvips]{graphicx}
\usepackage{bm,latexsym,amsmath,amssymb,amsfonts,color}
\usepackage[mathscr]{eucal}

\definecolor{mypink1}{rgb}{0.858, 0.188, 0.478}
\definecolor{mypink2}{RGB}{219, 48, 122}
\definecolor{mypink3}{cmyk}{0, 0.7808, 0.4429, 0.1412}
\definecolor{mygray}{gray}{0.6}

\newcommand{\be}[1]{\begin{equation} \label{#1}}
\newcommand{\ee}{\end{equation}}
\newcommand{\bea}{\begin{eqnarray}}
\newcommand{\eea}{\end{eqnarray}}
\newcommand{\ba}{\begin{array}}
\newcommand{\ea}{\end{array}}
\newcommand{\bel}{\begin{align}}
\newcommand{\eel}{\end{align}}

\newcommand{\tcr}{\textcolor{red}}

\newcommand{\tcgr}{\textcolor{mygray}}

\newcommand{\oo}{$\mathcal{O}$ }
\newcommand{\As}{$\overline{\mbox{AS}}$}

\begin{document}

\title{Black hole in closed spacetime with an anisotropic fluid}

\author{Hyeong-Chan Kim}
\email{hckim@ut.ac.kr}
\affiliation{School of Liberal Arts and Sciences, Korea National University of Transportation, Chungju 27469, Korea}
%
\begin{abstract}
We study spherically symmetric geometries made of anisotropic perfect fluid based on general relativity.  
The purpose of the work is to find and classify black hole solutions in closed spacetime.   
In a general setting, we find that a static and closed space exists only when the radial pressure is negative but its size is smaller than the density.  
The Einstein equation is eventually casted into a first order autonomous equation on two-dimensional plane of scale-invariant variables, which are equivalent to the Tolman-Oppenheimer-Volkoff (TOV) equation in general relativity.
Then, we display various solution curves numerically. 
An exact solution describing a black hole solution in a closed spacetime was known in Ref.~\cite{Cho:2016kpf}, which solution bears a naked singularity and negative energy era.
We find that the two deficits can be remedied when $\rho+3p_1>0$ and $\rho+p_1+2p_2< 0$, where the second violates the strong energy condition. 
\end{abstract}
\maketitle

\section{Introduction}
Even with the long observations, the spatial geometry of the Universe is an unresolved issue.
One reason is the (accelerating) expansion of the Universe, which restricts our observational boundary to a finite range inside a cosmological event horizon.
Even if one measures the average curvature of the observable Universe, it's topology cannot be answered definitely.
Philosophically, a closed space is favored because the total energy of an open Universe will have difficulty by definition. 
A closed space usually studied in a cosmological sense, therefore, time-dependent metrics are mainly used.
However, study about a static closed space may play important roles in the study of an early Universe and the dynamics of the Universe itself.
As shown in Ref.~\cite{Cho:2016kpf}, a static, homogeneous S$_3$ space is achieved only when $p= -\rho/3$, where $p$ and $\rho$ represent the pressure and the energy density, respectively.
For a constant $\rho$, time-dependent or space-dependent solution does not exist.

A space-dependent solution can be obtained when the density is allowed to be dependent on $r$. 
For example, a black hole solution inside a sphere was given by using $\chi \equiv \arcsin (r/R_0)$ coordinate,
\be{bh1}
ds^2 = - (1- K \cot \chi) dt^2 +
	\frac{R_0^2}{1- K \cot \chi} d\chi^2 + R_0^2 \sin ^2 \chi d\Omega_2^2 ,
\ee
where $K$ and $R_0$ are integrating constants. 
The energy density of the fluid is given by $\rho(\chi) = 3/8\pi R_0^2 \times (1-K \cot \chi)$.
On may easily find two deficits of the solutions.
First, there is a naked singularity on the opposite pole of the black hole.
Second, the energy density $\rho(\chi)$ may not be positive definite behind the event horizon.
Because of these, the solution~\eqref{bh1} cannot be used as a conceivable solution.
At the present work, we show that the two deficits can be cured when an anisotropic fluid is introduced and their equation of state is restricted appropriately.

A collection of static, spherically symmetric solutions of Einstein's field equation can be found in Stephani {\it et al}~\cite{Stephani2003}, Delgaty and Lake~\cite{Delgaty:1998uy}, and Semiz~\cite{Semiz:2008ny}.
Most of them focused on the isotropic fluids because astrophysical observations support isotropy.
Although the perfect Pascalian (isotropic) fluid assumption is supported by solid observational and theoretical grounds, an increasing amount of theoretical evidence strongly suggests that, for certain density ranges, a variety of very interesting physical phenomena may take place giving rise to local anisotropy (see \cite{Herrera1997} and references therein).
As investigated by Ruderman~\cite{Ruderman1972}, highly compact astrophysical objects having core density beyond the nuclear density ($\sim 10^{15} \mbox{g/cm}^3$) can have pressure anisotropy, i.e., the pressure inside these compact objects can be decomposed into two parts radial pressure $p_1$ and transverse pressure $p_2$ perpendicular direction to $p_1$. 
There are various reasons behind these anisotropic nature e.g., the existence of solid core in presence of type 3A superfluid~\cite{Kippenhahn1990}. 
Local anisotropy in self-gravitating systems were studied~\cite{Herrera1997,Bower1974,Mak:2001eb,Matese1980}.
The pressure anisotropy affects the physical properties, stability and structure of stellar matter~\cite{Dev2002}.
Recently, Bhar~\cite{Bhar2015,Ratanpal:2016kwu} studied a new model of an anisotropic superdense star which admits conformal motions in the presence of a quintessence field which is characterized by a parameter $w_q$ with $-1< w_q < -1/3$.
%
Charged anisotropic matter with linear equation of state~\cite{Thirukkanesh2008,Ivanov2002} and a self gravitating, charged and anisotropic fluid sphere~\cite{Varela2010,Bekenstein:1971ej} were also considered
For simplicity, we consider the static, spherically symmetric configurations.
The stress tensor for an anisotropic fluid compatible with the spherical symmetry is 
\be{st}
T_{\mu\nu} = (\rho + p_2) u_\mu u_\nu + (p_1 - p_2) x_\mu x_\nu + p_ 2 g_{\mu\nu},
\ee
where $\rho$ is the energy density measured by a comoving observer with the fluid, and $u^\mu$ and $x^\nu$ are its four-velocity and a spacelike unit vector orthogonal to $u^\mu$ and angular directions, respectively.
The radial and angular pressures are assumed to be proportional to the density:
\be{eos}
p_1 = w_1 \rho, \qquad p_2 = w_2 \rho.
\ee

We are interested in a static-spherically symmetric line element given by
\be{metric}
ds^2 = - f(r) dt^2 + g(r) dr^2 + r^2 d\theta^2 + r^2 \sin^2\theta d\phi^2.
\ee
We assume that the metric describes a spacetime having a blackhole horizon and an outer (event) horizon or outer maximal radius surface.
For the metric to have a correct signature, we need $f(r) g(r) > 0$.
We first assume that there is an apparent singularity of the metric at $r= a$.
Then, we assume the metric function around $r=a$ with the form  
\be{fg:a}
f(r) = (r-a)^\alpha \qquad g(r) = (r-a)^{-\beta},
\ee
where $\alpha$ and $\beta$ are real numbers.
Then, the first two diagonal components of the Einstein tensor  
$$
G^0_0 = r^{-2}\left(1+[-a +(1+\beta)r ](r-a)^{\beta-1} \right) 
\quad  G^1_1= r^{-2}\left(1+[-a +(1+\alpha)r ](r-a)^{\beta-1} \right) 
$$
are non-singular only if $\beta =0=\alpha$ or $\beta \geq 1$. 
The remaining nonvanishing component of the Einstein tensor,
$$
G^2_2 = \frac{(r-a)^{\beta-1}}{4r} \left[2\beta + \alpha(\alpha+\beta) \frac{r-2a/(\alpha+\beta)}{r-a}\right] ,
$$
may not be singular when $\beta \geq 2$ or $\beta \geq 1$ and $\alpha + \beta =2$. 
Unless $\beta$ is an integer, some components of higher derivatives of curvature will be singular. 
Therefore, for the geometry be non-singular at $r=a$, there are three possibilities:
\be{nonsingular}
\mbox{i)}\quad \beta = 1=\alpha, \qquad
\mbox{ii)}\quad \beta = 1,~ \alpha =0, \qquad
\mbox{iii)}\quad \beta = 2,3,4 \cdots.
\ee
For case iii), the geometry represents a cylindrical spacetime with maximum (or minimum) radius $a$.
For case i), the surface $r=a$ forms an event horizon. 
For case ii), the surface $r=a$ forms a (locally) maximal or minimal radius surface of S(3) geometry.
To have a blackhole spacetime in a closed spacetime, we require that there are an inner boundary and a maximal radius surface of the type i) and  ii), respectively.

In Sec. II, we reduce the Einstein equation into a first order autonomous equation on a two dimensional plane of dimensionless constants + a first order differential equation with respect to $r$. 
In Sec. III, the Einstein equation is solved around a point of coordinate singularity and find the conditions for the surface to be an event horizon and a maximal radius surface of a closed space.
In Sec. IV, the autonomous equation is classified with respect to the values of equation of state parameters $w_i$.
The behaviors of the solutions are analyzed also.
In Sec. V, we describe the exact solutions by means of the terminology of the present work to help the understanding of the readers.
In Sec. VI, we present various numerical solutions based on the classification of the solutions. 
We summarize and discuss the results in Sec. VII.

\section{Einstein equation}
The Einstein equation for the anisotropic fluid~\eqref{st} with the spherically symmetric metric in Eq.~\eqref{metric} becomes
\bea 
G^0_0 &=& -\frac1{r^2} + \frac{1}{r^2 g} - \frac{g'}{r g^2} = - 8\pi \rho(r), 
    \label{G00}\\
G^1_1 &=& -\frac1{r^2} + \frac{1}{r^2 g}+ \frac{f'}{rfg} = 8 \pi p_1(r),
	\label{G11} \\
G^2_2 &=& \frac{f'}{2rfg} - \frac{f'^2}{4f^2 g} - \frac{g'}{2rg^2} -
	\frac{f' g'}{4f g^2} + \frac{f''}{2fg} = 8\pi p_2(r), \label{G22}
\eea
where the prime denotes the derivative with respect to $r$. 
 
Equation~\eqref{G00} can be simplified by defining a mass function $m(r)$ as,   
\be{gr}
g(r) = \frac{1}{1-2m(r)/r}; \qquad m(r) = 4\pi \int^r r'^2 \rho(r') dr',
\ee
where an integration constant is absorbed into the definition of $M(r)$.
The Bianchi identity $\nabla^\mu T_{\mu\nu} =0$, combined with Eqs.~\eqref{gr} and \eqref{G11}, presents the ordinary TOV equation,  
\be{TOV}
p_1' = -(\rho + p_1) \frac{m+ 4\pi r^3 p_1}{r(r-2m)} + \frac{2(p_2 - p_1)}{r}.
\ee
With respect to $m(r)$, using $dm/dr = 4\pi r^ 2 \rho(r)$ and Eq.~\eqref{eos}, the TOV equation becomes, 
\be{TOV2}
\frac{m''}{m'} = -\frac{1+w_1}{w_1} \frac{ 1/2 
	+m'}{r-2m} +\frac{1+ w_1 + 4w_2}{2w_1 r},
\ee
where we use $\rho' /\rho = m''/m' - 2/r$.
Combining Eq.~\eqref{G00} with Eq.~\eqref{G11} and using the Bianchi identity, the $g_{tt}$ part of the metric can be integrated to give $f(r)$ from the density,
\be{f:rho}
f  \propto r^{4(w_1-w_2)/(1+w_1)} \rho^{-2w_1/(1+w_1)} .
\ee
Equation~\eqref{G11} can also be directly integrated to give 
\be{fr:g}
 f(r)= \frac{f_0}{r}  \left[\frac{g(r)}{r}\right]^{w_1}\exp \left[(1+w_1) \int \frac{g(r)}{r} dr\right]  
 	=  \frac{f_0(r-2m)^{-w_1}}{r}  \exp \left[(1+w_1) \int \frac{1}{r- 2m(r)} dr\right].
\ee
Note that the Einstein equation does not determine the sign of $f(r)$.
Therefore, the value of $f_0$ will be chosen for convenience so that the signature of the metric be Lorentzian.
The remaining task is to find the explicit functional dependence of $m(r)$ by solving the TOV equation. 
When $w_1=-1$ and $w_1=-1/3=w_2$, the TOV equation allows exact solutions as in Ref.~\cite{Cho:2017nhx,Cho:2016kpf}. 

For other cases, it is not easy to find an analytic solution.
However, numerical solutions are  still available. 
Let us simplify the equation after introducing scale invariant variables 
\be{uv}
u \equiv \frac{2m(r)}{r}, \qquad v \equiv \frac{dm(r)}{dr} =4\pi r^2 \rho,
\ee
and logarithmic change of radius $\xi$ as
\be{xi}
 \xi = \log \frac{r}{r_+},
\ee
where $r_+$ is a to be determined scale of radius.
Now, the radius change can be written by using $u$ and $v$ as
\be{dxi}
d\xi = \frac{du}{2v-u} , \qquad v \equiv \frac{dm(r)}{dr} = 4\pi r^2 \rho = \frac{u+u'}{2},
\ee
where, from now on in this work, the prime represents the derivative with respect to $\xi$.
Notice that, as $u$ increases, the radius increases/decreases when $u \lessgtr 2v$. 
In terms of the scale invariant variables, the TOV equation~\eqref{TOV2} can be rewritten as an autonomous equation of the form
\be{de2}
\frac{du}{dv} 
 = \frac{-1}{1+w_1} \frac{(1-u)(2v-u)}{v\left[ v- v_M
 	+ s (1-u)\right]},
\ee
where $v_M$ and $s$ represent
\be{slope}
v_M \equiv -\frac{1}{2w_1}, \qquad s \equiv \frac{1+w_1+ 4w_2}{-2w_1 (1+ w_1)}.
\ee
We call the integral curve of Eq.~\eqref{de2} a solution curve $C$.
When $ w_2 = -(1+w_1)/2$, i.e., $s= 1/2w_1$, Eq.~\eqref{de2} allows a linear solution curve
\be{linear sol}
v = v_M u .
\ee
The linear solution plays an important role in analyzing the solution space of Eq.~\eqref{de2}.
Note that the condition $w_2 = -(1+w_1)/2$ is the boundary of the strong energy condition $\rho + \sum_i p_i\geq 0$.

Given $u$ and $v$ as functions of $r$, the metric functions are given by, from Eqs.~\eqref{gr} and \eqref{f:rho},
\be {fg:uv}
f(r) =  f_0\, r^{4(2w_1-w_2)/(1+w_1)} \left| v\right|^{-2w_1/(1+w_1)} ,  \qquad 
g(r) = \frac{1}{1-u}, 
\ee
where the signature of $f_0$ should be determined so that the metric is Lorentzian. 
Note that the signatures of $g(r)$ changes across the line $u=1$.
Therefore, $r$ plays the role of a space/time coordinate for $u\lessgtr 1$. 
Let us assume a solution curve $C$ crosses the line $u=1$ through a point P.
If this happens, the signature of $f(r)$ also changes across this line to keep the Lorentzian signature, which implies the appearance of a discontinuity of the metric function $f(r)$. 
To avoid the discontinuity, the value $f(r)$ should goes to zero when $C$ crosses $u=1$, which implies $v\to 0$ there.  
Therefore, any (physically relevant) solution curve crosses $u=1$ only through the point $(u,v) = (1,0)$.

\section{Inner horizon and maximal radius surface}
Let $r=r_-$ forms an inner boundary of a static region of a spherically symmetric spacetime described by the metric~\eqref{de2}.
The boundary will form an event horizon. 
Because there is a coordinate singularity, we first write $g(r) \approx g_- (1-r_-/r)^{-1}$ and find $f(r)$ from Eq.~\eqref{fr:g}, to get
$$
f(r)  
 \propto \frac{g_-^{w_1} }{r} (r-r_-)^{(1+w_1)g_- -w_1}  .
$$
Because $r=r_-$ is an event horizon, $f(r) \propto 1-r_-/r$ and $\displaystyle \lim_{r\to r_-} f(r)g(r)> 0$ should be satisfied. 
Therefore, $r_-$ forms an event horizon only when
\be{inner_H}
 w_1=-1 ~~\mbox{ or }~~ g_-=1.
\ee
Behind the horizon, $r$ plays the role of time. 
Therefore, $-p_1$ and $-\rho$ play the role of energy density and spatial pressure for $r< r_-$, respectively.
At the horizon, in the absence of a singular surface density, the energy density and pressure must be continuous. 
This requires $\displaystyle \lim_{r\to r_-+0}\rho(r) = - \lim_{r\to r_--0}p_1(r)$. 
There are two ways satisfying this.
i) $w_1 = -1$. ii) $\rho(r_-) = 0$. 
As will be shown soon, the value of $w_1$ cannot satisfy $w_1 = -1$ if there is a maximal radius surface so that the space is static and closed.  
Disregarding this case i), the density must vanish at the horizon to form an event horizon, which gives $v \to 0$ once more.

Let $r=r_+$ be a surface of (locally) maximal radius of a S$_3$ geometry described by the metric~\eqref{metric}.
Let us write $g(r) \approx g_+ (r_+/r-1)^{-1}$ with $g_+ > 0$ and calculate Eq.~\eqref{fr:g} around $r_+$ to get
\be{outer}
f(r)\approx  \frac{f_0}{r}\Big(\frac{g(r)}{r}\Big)^{w_1}  
	\exp \left[ -(1+w_1)g_+ \int^r \frac{1}{ r-r_+} dr\right]
\propto \frac{g_+^{w_1} }{r} (r_+ -r)^{-w_1-(1+w_1)g_+}  .
\ee
Because $r_+$ is a maximum of the radius in a S$_3$ geometry, the metric function $f(r)$ should approach a non-divergent positive number at $r=r_+$.
This determines 
\be{outer}
g_+ = -\frac{w_1}{1+w_1} .
\ee
Note that $w_1$ should satisfy $-1< w_1 < 0$ because $g_+$ is a positive number. 
When $w_1 = -1$, the outer boundary cannot form a surface of maximal radius.
Around the maximal value of $r$, $m(r)$ behaves as 
$$
m(r)  \simeq \frac{1+g_+}{g_+} r - \frac{r_+}{g_+}.
$$
This gives 
\be{v:vm}
v = \frac{dm}{dr} \simeq \frac{1+g_+}{2g_+} =v_M,
\ee
where we uses the explicit value of $g_+$ in Eq.~\eqref{outer} at the last equality.
Note that the size of the maximal value of the radius is determined by the value of the radial pressure and is independent of the angular pressure.

\vspace{1cm}

\section{Analysis for general solution}

An integral curve $C$ to the autonomous equation~\eqref{de2} can be plotted on a two dimensional surface $(u,v)$. 
At the present work, we restrict our interest to the case $-1< w_1 < 0$ because we are interested in solutions whose spatial topology is S$_3$.

There are four interesting lines which determine properties of $C$.
On the first two lines
\be{Rline}
\mbox{R1:} ~v=0, \qquad \mbox{R2:} ~ v =s u + \frac{2 w_2}{w_1(1+w_1)}
\ee
 only the value $u$ changes, i.e. $\delta v=0$.
On the first line R1, the energy density $\rho$ goes to zero.
On the other two lines
\be{Bline}
\mbox{B1:}~ u=1, \qquad \mbox{B2:} ~u=2v
\ee 
only the value of $v$ changes, i.e. $\delta u=0$. 
The line B1 represents a static boundary given by $r = 2m(r)$.  
Because the line B1 is parallel to the $v$ axis, $C$ cannot cross the line.
Similarly, R1 is parallel to the $u$ axis. 
Therefore, $C$ cannot cross the line too.
In fact, the crossing can occur only through the points where the R lines meet the B lines, which are
 \be{crossing}
\mbox{P1:} ~ (u,v) = (1,0), \qquad \mbox{P2:} ~ (u,v) = \big(1, v_M \big), \qquad
\mbox{P3:}~ (u,v) = (2v_3, v_3); \quad v_3\equiv \frac{2w_2}{(1+w_1)^2+ 4w_2}.
\ee
Especially, $C$ can pass through the line B1 only through the points P1 and P2.

\begin{figure}[htb]
\begin{center}
\begin{tabular}{ccc}
\includegraphics[width=.3\linewidth,origin=tl]{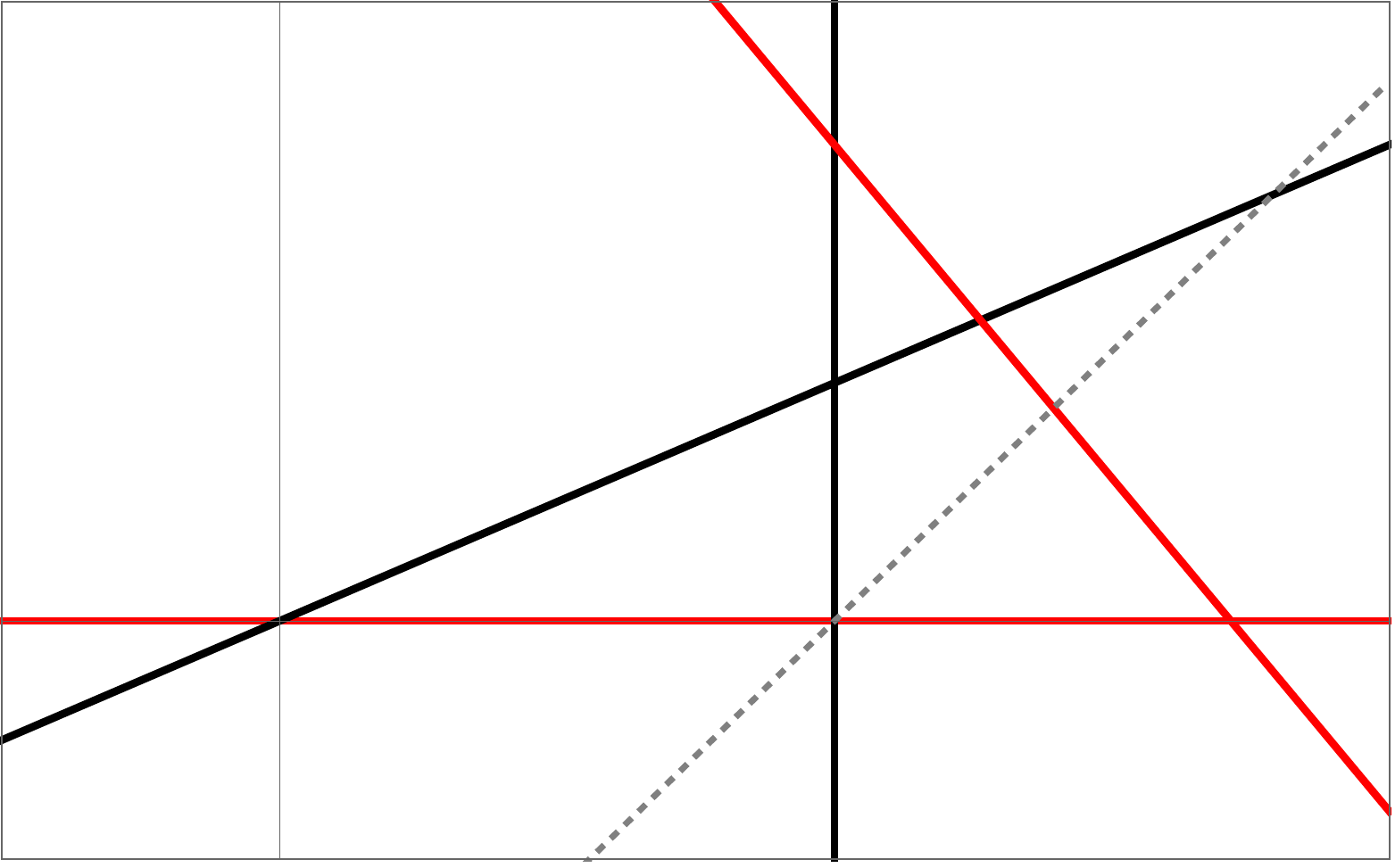}
&
\quad
\includegraphics[width=.3\linewidth,origin=tl]{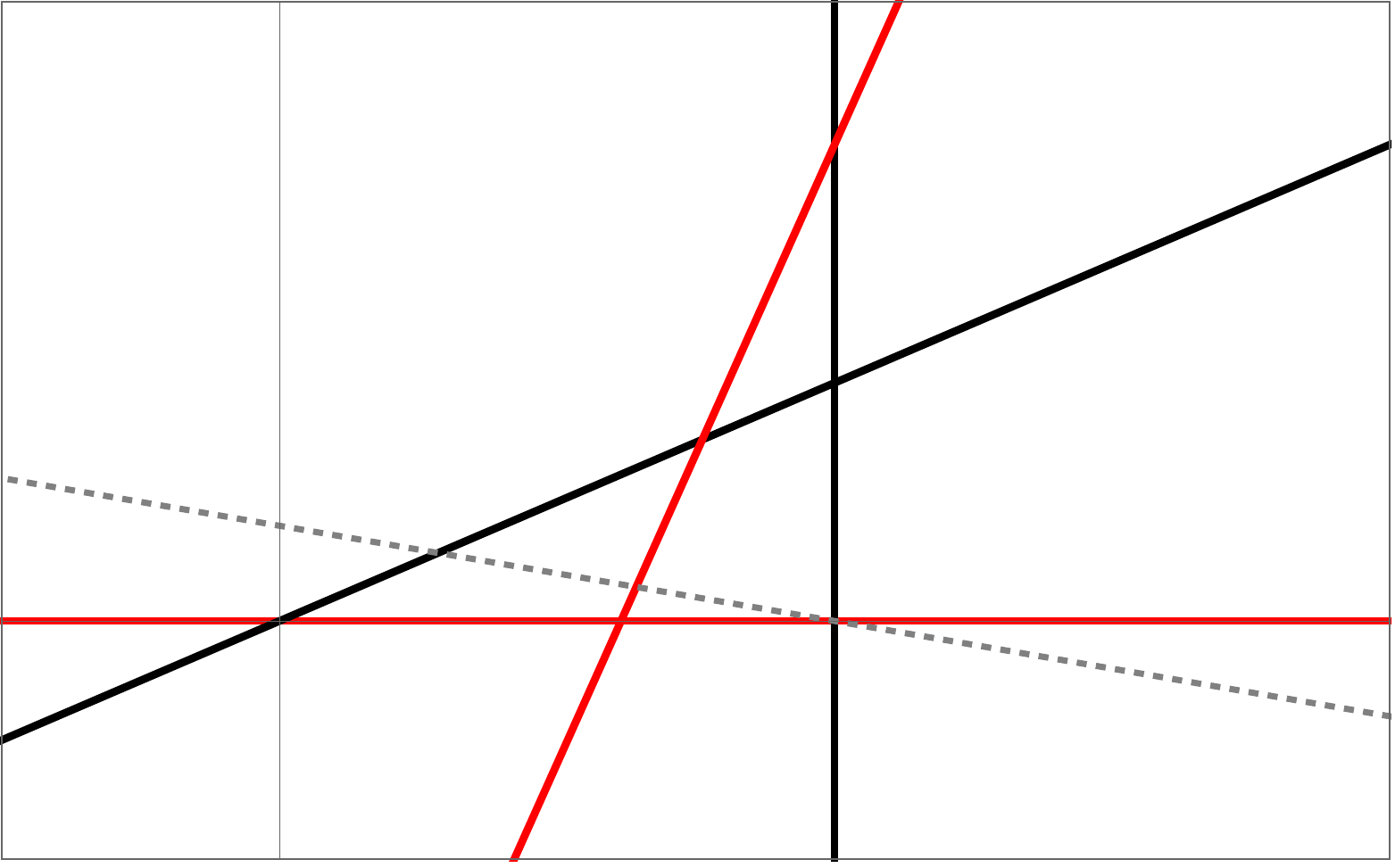}
&
\quad
\includegraphics[width=.3\linewidth,origin=tl]{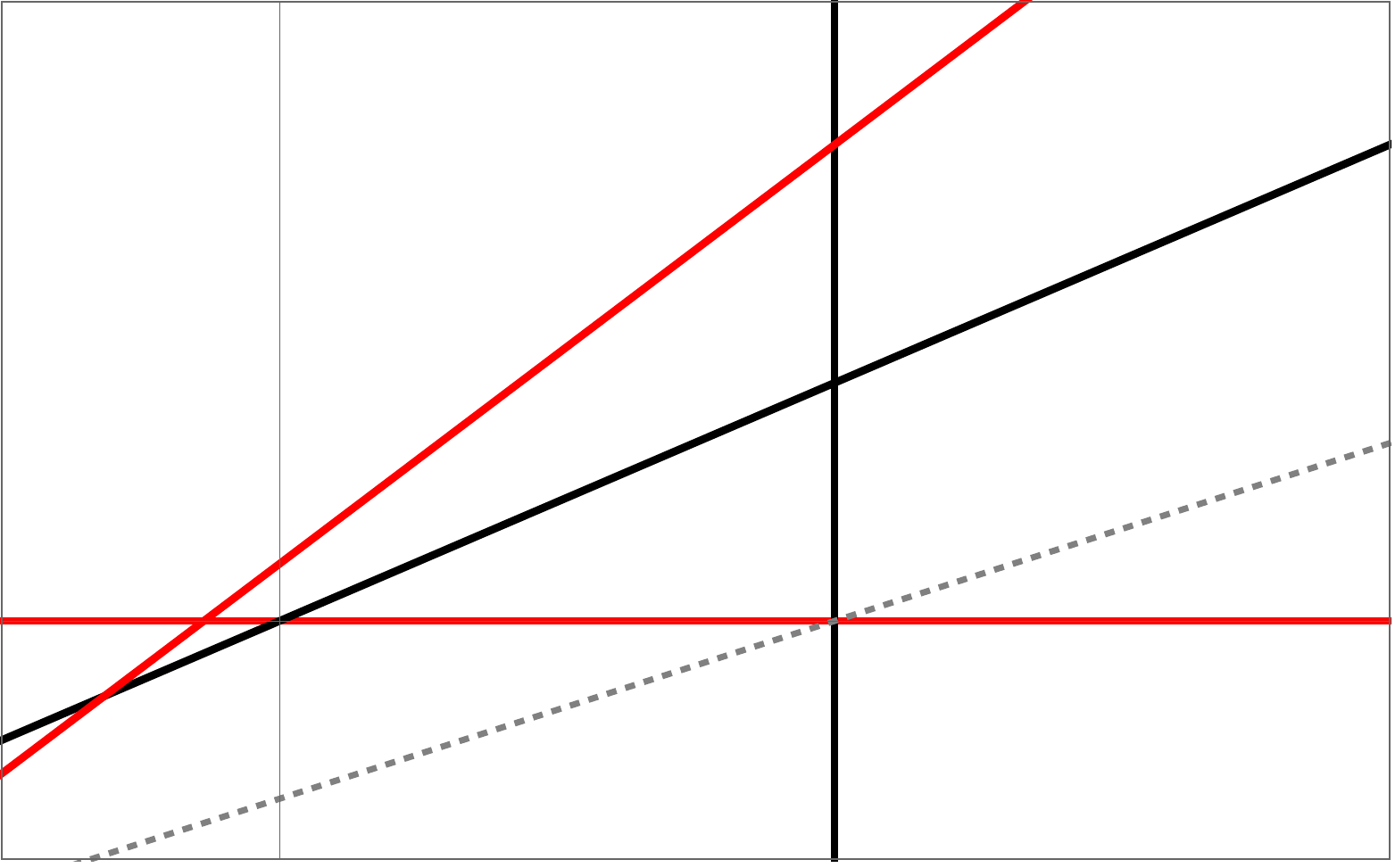}
\\
Type I & Type II & Type III
\end{tabular}
\put (-334,-14) {$u$  }
\put (-460,60) {$v$  }
\put (-460,-20) { $\mathcal{O}$  }
\put (-350, 40) {B2}
\tcr{\put (-485, -10) {R1}
\put (-420, 55) {R2}}
\put (-395, -30) {B1}
\tcgr{\put (-420, -30) {\As}}
\put (-396,38) { P2} \put (-399,38) {\tcr{\large $\bullet$}}
\put (-409,-20) { P1} \put (-399,-15) {\tcr{\large $\bullet$}}
\put (-388,11) { P3} \put (-383,18) {\tcr{\large $\bullet$}}
\put (-410,15) {\footnotesize{ 0.5} }
\put (-380, 35) {\textcolor{cyan}{ \Huge $\rightarrow $  }}
\put (-365, 12) {\textcolor{cyan}{ \Huge $\leftarrow $  }}
%
\put (-168,-14) {$u$  }
\put (-294,60) {$v$  }
\put (-294,-20) { $\mathcal{O}$  }
\put (-184, 40) {B2}
\tcr{\put (-319, -10) {R1}
\put (-259, -35) {R2}}
\put (-229, -30) {B1}
\tcgr{\put (-319, 5) {\As}}
\put (-230,38) { P2} \put (-233,38) {\tcr{\large $\bullet$}}
\put (-243,-20) { P1} \put (-233,-15) {\tcr{\large $\bullet$}}
\put (-259,11) { P3} \put (-247,5) {\tcr{\large $\bullet$}}
\put (-232,10) {\footnotesize{ 0.5} }
\put (-214, 35) {\textcolor{cyan}{ \Huge $\rightarrow $  }}
\put (-199, 12) {\textcolor{cyan}{ \Huge $\leftarrow $  }}
%
\put (0,-14) {$u$  }
\put (-128,60) {$v$  }
\put (-128,-20) { $\mathcal{O}$  }
\put (-18, 40) {B2}
\tcr{\put (-153, -10) {R1}
\put (-93, 28) {R2}}
\put (-63, -35) {B1}
\tcgr{\put (-19, 5) {\As}}
\put (-64,38) { P2} \put (-67,38) {\tcr{\large $\bullet$}}
\put (-67,-20) { P1} \put (-67,-15) {\tcr{\large $\bullet$}}
\put (-147,-28) { P3} \put (-146,-22.5) {\tcr{\large $\bullet$}}
\put (-66,10) {\footnotesize{ 0.5} }
\put (-48, 35) {\textcolor{cyan}{ \Huge $\rightarrow $  }}
\put (-33, 12) {\textcolor{cyan}{ \Huge $\leftarrow $  }}
\end{center}
\caption{
Classification of the autonomous equation. 
The cyan arrows represent the increasing direction of radial coordinate $r$.  
The direction changes based on the line B2.
The red line R2 changes depending on the values of $w_i$.
In this figure, we choose $w_1 = -1/2$, and $w_2 =-0.3 , ~ 0.2,$ and $ -0.015$, respectively for cases I, II, and III.
The character $\mathcal{O}$ represents the point $(u,v) = (0,0)$.
The dotted gray line denoted by the character `\As' is the asymptotic line in Eq.~\eqref{charline}.
}
\label{fig:cases}
\end{figure}
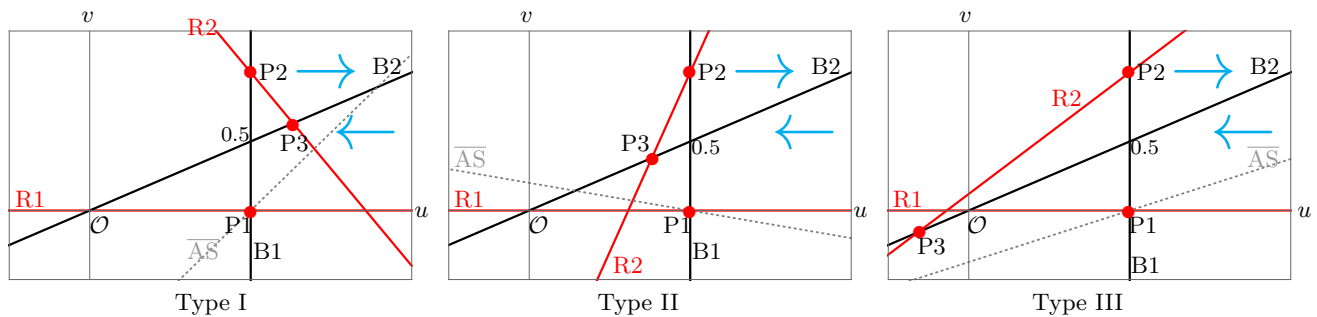

By means of the slope $s$, we divide the deployment of the points and the lines into three different types as in Fig.~\ref{fig:cases},
\be{cases}
\mbox{I:} ~s< \frac{1}2 ~~( w_2 < -\frac{(w_1+1)^2}{4} ), \qquad
\mbox{II:}~  s \geq v_M ~~ ( w_2 \geq 0 ), \qquad
\mbox{III:}~  \frac12 \leq s < v_M~~ ( -\frac{(w_1+1)^2}{4} \leq w_2 < 0). 
\ee
For the types I, II, and III, the point P3 is located in the region $(u> 1, v> 0)$, $(0< u< 1, v> 0)$, and $(u< 0, v<0)$, respectively. 
The point P2 in Eq.~\eqref{crossing}, which represents a maximal radius surface, is located upper than the line B2 because $|w_1|<1$.
$C$ crosses the B and the R lines vertically and horizontally, respectively.
%
One can catch various asymptotic properties from Fig.~\ref{fig:cases} presents.
For example, for the case of type I, the limit $(u\to \infty, v)$ corresponds to a beginning of spacetime because the coordinate $r$ is timelike and $r$ coordinate increases as $u$ decreases in the region below B2.
The limit $(u\to -\infty, v)$ corresponds to a naked singularity with $r=0$ because $r$ is a space coordinate and its size will be monotonically decreases as $u$ decreases. 
The limit $(u\to \infty, v> u/2)$ corresponds to a future infinity because the time $r$ increases as $u$ increases in the region.

\subsubsection{ Behavior of a solution around the point $\mathcal{O}$ } \label{sec:O}
Let us first search for the behavior of $C$ around the point $\mathcal{O}$, given by $(u,v)= (0,0)$.
We put the trial function $u = \kappa v^\beta$ with $|v| \ll 1$ and $\beta > 0$ to Eq.~\eqref{de2} and find that there are two possibilities: 
\begin{eqnarray} \label{origin}
\mbox{i):}& u = \frac{2w_1}{w_1 + 2w_2} v, & ~\\
\mbox{ii):} & u = \kappa  v^{-w_1/2w_2}, &\qquad w_2> \frac{-w_1}{2} >0,      \label{origin2}
\end{eqnarray}
where the first shows a linear behavior and the second shows a polynomial behavior.
The polynomial behavior happens only for the type II case in Eq.~\eqref{cases} because $w_2> -w_1/2$. 
For both of the possibilities, the radius behaves as 
$$
r = r_0 v^{w_1/2w_2} .
$$
Therefore, $\mathcal{O}$ represent the center of the star $(r=0)$ when $w_1$ and $w_2$ are of the same signature. 
On the other hand,  $r\to \infty$ when $w_1$ has opposite signature to $w_2$.
The density takes the form,  
\be{rho:origin}
\rho = \frac{v}{4\pi r^2} =\frac{1}{4\pi r_0^{2w_2/w_1}} \frac1{ r^{2+2w_2/(-w_1)} }.
\ee
When $w_2> -w_1/2$, the density decreases faster than $r^{-3}$ as $r\to \infty$. 
When $0< w_2< -w_1/2$, the density decreases to zero as $r \to \infty$, however, its integration over the space diverges. 
When $-w_1< w_2<0$, the density diverges at the origin as $r\to 0$. 
When $w_2< w_1<0$, the density go to zero as $r\to 0$.

Summarizing, there are four different behaviors around the point $(0,0)$, which are
\be{Origin:behave}
\begin{aligned}
\mbox{O1:} &\quad r= r_0 v^{w_1/2w_2} \approx 0, &  w_2 \leq w_1, & \qquad u = \frac{2w_1}{w_1+2w_2}v , \quad \rho \to \rho_0 \\
\mbox{O2:} &\quad r= r_0 v^{w_1/2w_2} \approx 0, &  w_1< w_2< 0, & \qquad u = \frac{2w_1}{w_1+2w_2}v , \quad \rho \to \infty \\
\mbox{A1:}  & \quad r= r_0 v^{w_1/2w_2} \to \infty, &
	w_2>0,   & \qquad   u = \frac{2w_1}{w_1+2w_2}v, \quad \rho \to 0 \\
\mbox{A2:}  & \quad r= r_0 v^{w_1/2w_2} \to \infty, &
	w_2> \frac{-w_1}{2},   & \qquad   u = \kappa v^{-w_1/2w_2} , \quad \rho \to 0 
\end{aligned},
\ee
where in the first line, $\rho_0 =0$ when $w_2 < w_1$. 
The characters `O' and `A' represent that the behavior happens at the center of the star and in the asymptotic region, respectively. 
Note that both asymptotic behaviors can happen at \oo when $w_2> -w_1/2$. 

\subsubsection{Behavior of a solution around {\rm P1:}  the event horizon}
Let us search for the behavior of $C$ around P1$(1,0)$.
We put the trial function $1-u= \kappa |v|^\beta$ with $\beta>0$ to Eq.~\eqref{de2}. 
Then, we get 
\be{SC:P1} 
 \beta = \frac{-2w_1}{1+w_1} \quad \Rightarrow \quad 1-u = \kappa |v|^{-2w_1/(1+w_1)}.  
\ee
The radius can be obtained by integrating $d\xi = du/(2v-u) \approx -du$ to get 
\be{r:P1}
 r = r_- e^\xi \approx r_- e^{1-u}. 
\ee 
Therefore, the radius smoothly changes with respect to $u$.
The density around $r_-$ is $\rho \sim v/(4\pi r^2) \approx 0$. 

For $\beta > 1$ ($-1<w_1< -1/3$), a differentiable solution curve passes P1 vertically, parallel to the $v$ axis. 
%
For $0< \beta < 1$ ($-1/3< w_1< 0$), $C$ passes P1 horizontally, parallel to the $u$ axis.
For $\beta = 1$ ($w_1 = -1/3$), $C$ passes the point P1 with a non-zero finite slope. 
Only for the case with $-1/3< w_1< 0$, there exists a differentiable solution curve which passes P1 and has a non-negative energy density ($v\geq 0)$ at the both sides of P1.

\subsubsection{ Behavior of a solution around  {\rm P2:} the maximal/minimal radius surface}
Let us search for the behavior of $C$ around P2$(1,v_M)$.
By using the trial function $1-u = \kappa |v-v_M|^\beta$, we find that Eq.~\eqref{de2} allows the quadratic and the linear behaviors for $C$,
\begin{eqnarray} \label{hor}
\mbox{i)}&& ~\beta = 2,   \qquad  u =1+ \kappa  (v-v_M)^2, \\
\mbox{ii)} &&~\beta =1,   \qquad u = 1 - s^{-1} ( v- v_M),  \label{hor2} 
\end{eqnarray}
where $\kappa$ represents an arbitrary real number and $s$ is given in Eq.~\eqref{slope}.

For the case i), the radius takes the form,  
\be{r:P2_i}
r \approx r_0 \left[1+ \frac{-w_1\kappa}{1+w_1}(v-v_M)^2 \right].
\ee
Therefore, $r$ takes minimum/maximum value at $v= v_M$ when $\kappa \gtrless 0$.  
For $u\leq 1$, $r_0$ plays the role of a maximum value of the radius.
For $u \geq 1$, the time-dependent scale $\sqrt{g_{tt} g_{\theta\theta}g_{\phi\phi}}$ bounces at P2 and has a minimum value at $r_0$.  
Therefore, P2 presents an extremum of the radial coordinate $r$.

For the case ii), the radius becomes linear in $v$ as $\xi \approx \xi_0 + \xi_0' v$ with $\xi_0' = -2w_1^2/(1+w_1 + 4w_2)$. 
The metric component $g_{rr} = g(r) =(1-u)^{-1}$ changes signature at $u=1$. 
On the other hand, $g_{tt}$ goes to a finite value as $(u,v) \to {\rm P2}$. 
To keep the Lorentzian signature without a singularity at P2, a solution curve of type ii) is not allowed to pass P2.
In this sense, as will be seen in Figs.~\ref{fig:iso}-\ref{fig:caseIII}, the case ii) can be understood as a large $\kappa$ limit of the case i).

\subsubsection{Behavior of a solution around {\rm P3:} the bouncing point}
Around the point P3, after setting $u = u_3 + x$ and $v= v_3 + y$,  the differential equations~\eqref{dxi} and~\eqref{de2} can be written as to the linear order in $x$ and $y$,
\be{dxy/xi}
\frac{d}{d\xi}  \left( \begin{array}{c}
    x \\ 
    y \\ 
  \end{array} \right)=   \left(\begin{array}{cc}
    -1 & 2 \\ 
    c &-cs^{-1} \\ 
  \end{array}  \right)\left( \begin{array}{c}
    x \\ 
    y \\ 
  \end{array} \right) , \qquad  
 c \equiv \frac{w_2(1+w_1+4w_2)}{-w_1(1+w_1)^2} .
\ee
Defining a new variables $X_\pm$ as 
$$
X_+ = - \frac{s-c + \sqrt{8cs^2+(s-c)^2}}{2sc} x + y, \qquad
X_-  = - \frac{s-c - \sqrt{8cs^2+(s-c)^2}}{2sc} x + y,
$$
the equation~\eqref{dxy/xi} becomes
\be{X:asym}
\frac{dX_\pm}{d\xi} = \epsilon_\pm X_\pm; 
\qquad \epsilon_\pm =-\frac{(c+s)  \pm \sqrt{8cs^2+(s-c)^2}}{2s}.
\ee
The solution is $X_\pm = X_{\pm,0}\, e^{\epsilon_\pm \xi}.$
Note that the size of the squared-root term is larger than $|c+s|$ for types I and II because 
$$
8cs^2+(s-c)^2 -(c+s)^2 = 4cs(2s-1) = \frac{8s^2}{-w_1(1+w_1)^2} w_2[4w_2 +(w_1+1)^2]> 0.
$$
This implies that one of the eigenvalues is positive and the other is negative.
Therefore, every solution curve will follow in along one of the $X_\pm$ and follow out along the $X_\mp$. 
For the type III, the two eigenvalues have the same signature. 
However, we are not interested in this case because P3 is located in negative energy region. 

\vspace{.3cm}
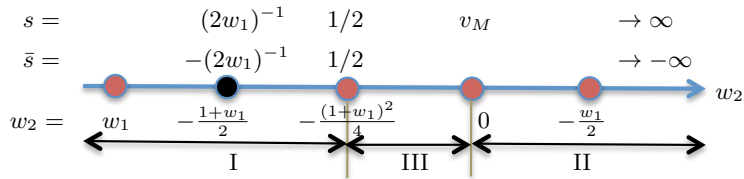
\begin{figure}[bht]
\begin{center}
\begin{tabular}{c}
\includegraphics[width=.5\linewidth,origin=tl]{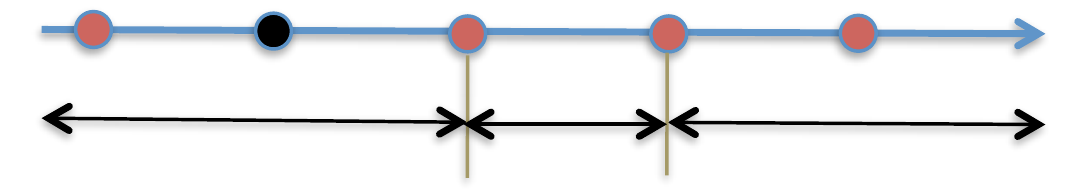} 
\end{tabular}
\put (-270, 27) {$\bar s =$} \put(-270, 42) {$s =$} \put(-275,4) {$ w_2 = $}
\put (-240,4){$w_1$}
\put (-212,4){$-\frac{1+w_1}{2}$} \put (-203,42) {$ (2w_1)^{-1}$}  \put(-209,27) {$ -(2w_1)^{-1} $}
\put (-155,27){$1/2$} \put (-155,42) {$ 1/2$}  \put(-166,4) {$ -\frac{(1+w_1)^2}{4} $}
\put (-98, 4) {$0$} \put (-105, 42) {$v_M$}
\put (-68, 4) {$ -\frac{w_1}{2}$}
\put (-45, 42) {$ \to \infty$}  \put (-45, 27) {$ \to -\infty$}
\put (-8, 15) {$w_2$}
\put (-192,-12) {I}  \put(-65,-12){ II} \put (-130,-12) { III}
\end{center}
\caption{Important values of $w_2$, $s$ and $\bar s$ for the classification of the behaviors around $\mathcal{O}$, P1, and P2.  
}
\label{fig:pts}
\end{figure}
The behaviors of a solution curve around the points $\mathcal{O}$, P1 and P2 are summarized in table~\ref{table1} and various important values are compared in Fig.~\ref{fig:pts} on a straight line.
The value $w_2 = -(1+w_1)/2$, which presents an exact solution curve in Eq.~\eqref{linear sol}, is also shown.
The values of $s$ and $\bar s$ for each points are displayed too, where $\bar s$ will be defined in Eq.~\eqref{u:-inf}.
\begin{table}[htp]
\caption{The behavior of a solution curve around $\mathcal{O}$, P1, and P2}
\begin{center}
\begin{tabular}{|c|c|c|c|c|c|}
\hline 
	Role  & 
	Position in $(u,v)$ &
	Condition & 
	$r/r_0$ & 
	$u-v$ & 
	$\rho$ 
	\\
\hline
     O1 & $\mathcal{O}$ & 
		$w_2 \leq w_1< 0$& 
		$ v^{w_1/2w_2}$ & 
		$u= \kappa v$ &
		$ r^{-2+ 2w_2/w_1} \to \rho_0$ \\	
\hline
   O2 & $\mathcal{O}$ &
	$w_1< w_2 < 0$ & 
	$v^{w_1/2w_2}$ & 
	$ u= \kappa v$ &  
	$r^{-2+2w_2/w_1} \to \infty$  \\
\hline
  A1 &
  	$\mathcal{O}$ &
	$w_2> 0$ &
	$ v^{w_1/2w_2}$ &
	$u =\kappa v $ &
	$r^{-2+2w_2/w_1} \to 0 $ \\
\hline
  A2 &
  	$\mathcal{O}$ &
	$w_2> -w_1/2$ &
	$ v^{w_1/2w_2}$ &
	$u \propto v^{-w_1/2w_2} $ &
	$r^{-2+2w_2/w_1} \to 0 $ \\
\hline
  H1 &
  	P1, vertical &
	$-1 < w_1 <-1/3$ &
	$r_H e^{1-u}$ &
	$1-u \propto |v|^{-2w_1/(1+w_1)} $ &
	$v/4\pi r^2 \approx 0 $ \\
\hline
  H2 &
  	P1, linear &
	$ w_1 = -1/3$ &
	$r_H e^{1-u}$ &
	$1-u \propto v $ &
	$v/4\pi r^2 \approx 0 $ \\	
\hline
  H3 &
  	P1, horizontal &
	$-1/3 < w_1<0$ &
	$r_H e^{1-u}$ &
	$1-u \propto |v|^{-2w_1/(1+w_1)} $ &
	$v/4\pi r^2 \approx 0 $ \\
\hline
  M1 &
  	P2 $(1,v_M)$ &
	~ &
	$r_0\pm\kappa'(v-v_M)^2$ &
	$|1-u| \propto |v-v_M|^{2} $ &
	$\rho_1+O(\delta v^2) $ \\
\hline
  M2 &
  	P2 $(1,v_M)$ &
	~ &
	$r_0+\kappa'(v-v_M)$ &
	$1-u =s^{-1} (v-v_M)$ &
	$\rho_1+O(\delta v) $ \\
\hline
\end{tabular}
\end{center}
\label{table1}
\end{table}%
The characters `O', `A', `H', and `M' represent that the position plays the roles of an origin ($r=0$), of an asymptotic region ($r\to \infty$), of an event horizon, and of a maximal (minimal) $r$ surface, respectively.
$\rho_0$ and $\rho_1$ represent a given value zero or not and a given non-vanishing value, respectively. 
$\kappa = 2w_1/(w_1+ 2w_2)$.
$r_0$ and $\kappa'$ are the maximal (minimal) value of radius and an appropriate constants representing how to approach the extremum, respectively.

\subsubsection{ Behavior of a solution in the large limit  $ |1-u|$ and/or $ |v| $ } \label{sec:asym}
Introducing $g(v) \equiv [1-u(v)]/v$ after ignoring constants compared to $u$ and $v$, the equation of motion~\eqref{de2} becomes 
\begin{eqnarray} \label{u:-inf}
\frac{1+ sg}{1+ \bar s g} \frac{dg}{g} 
=  \frac{1-w_1}{1+w_1} \frac{dv}{v} ;
	 \qquad \bar s \equiv \frac{1+3w_1 + 4w_2}{2w_1(1-w_1)} .
\end{eqnarray}
The solution to this is  
\bea \label{g:v}
 \log |g| - \left(1-\frac{s}{\bar s}\right) \log \left|1+ \bar s g\right| = \frac{1-w_1}{1+w_1} \log \left|\frac{v}{v_0}\right|,
\eea 
where $v_0$ is an integration constant. 
Because $|v/v_0| \gg 1$ or $|u|\gg 1$, Eq.~\eqref{g:v} presents three distinct limits, 
\bea
\mbox{A:} & \quad  &|g| \ll 1 \quad\quad ~\Rightarrow  \quad  \frac{v}{v_0} = |g|^{(1+w_1)/(1-w_1)}, \qquad 
|g|= \left|\frac{1- u}{v}\right| \approx \left(\frac{v}{v_0} \right)^{(1-w_1)/(1+w_1)}  , \label{g:lim1}\\
\mbox{B:} & \quad & |g| \gg 1 \quad \quad ~\Rightarrow \quad |g| 
	= \left|\frac{1- u}{v}\right| \approx  \left|\frac{v}{v_0}\right|^{-(1+ 3w_1+ 4w_2)/(1+w_1+ 4w_2)} , \label{g:lim2} \\
\mbox{C:} & \quad & g \to -\bar s^{-1} \quad \Rightarrow \quad 
	g = \frac{1-u}{v}
   \approx -\bar s^{-1}\Big[1 \pm \left(\frac{v}{v_0}\right)^{    
	-\frac{(1-w_1)(1+ 3w_1 + 4w_2)}{2[(1+w_1)^2 + 4w_2]} } \Big] 
	. \label{g:lim3}
\eea
The limit A is achieved in the region $v \gg |u|$ when $|w_1|>1$.
This case is out of the interests of the present work. \\
The limit B [the large $g$ limit] is achieved in the region $|u| \gg 1$ and $|v/v_0| \gtrless 1 $ when $[ -(1+ w_1)/4< w_2< - (1+3w_1)/4]$/$ [w_2> -(1+ 3w_1)/4$ or $ w_2< - (1+w_1)/4$], respectively.
The solution curve satisfies 
\be{g:lim22}
v \approx v_0  \left|\frac{1-u}{v_0}\right|^{(1+w_1+ 4w_2)/(-2w_1)}. 
\ee 
Integrating Eq.~\eqref{dxi} by using $|u| \gg |v|$, the radius and the density behave as 
\be{rho:lim2}
r \approx \frac{r_0}{|u|} \to 0, \qquad 
\rho \approx \frac{v_0^3}{4\pi r_0^2} \left( \frac{v_0 r}{r_0} \right)^{-(1- 3w_1+ 4w_2)/(-2w_1)}.
\ee
Therefore, this limit describes behaviors around the origin. 
The density diverges/vanishes at the origin when $w_2 \gtrless (3w_1-1)/4$. 
The explicit behaviors are denoted as O3-O6 in Table~\ref{table2}.
~\\
The limit C is achieved when a solution curve is located around an asymptotic line of slope $\bar s$ given by
\be{charline}
v = \bar s (u-1) ,
\ee
which is plotted as a gray dotted line in Fig.~\ref{fig:cases}.
Along the line, the sizes of $u$ and $v$ increase indefinitely. 
For this behavior happens, the correction term in the square bracket in the right hand side of Eq.~\eqref{g:lim3}  should goes to zero in the limit. 
Because the size of $v$ is large, the exponent of the correction term should be negative.
This determines the limit exists only for the types I [$\bar s> 1/2$ and $s< 1/2$] and II [$\bar s \leq 0$ and $s \geq v_M$].
For the type I, the slope $\bar s$ is positive definite. 
On the other hand, for the type II, $\bar s$ is negative definite.
The radius and the density behave as
\be{r rho:A5}
r \approx r_0\left( \frac{u_0}{u}\right) ^{ 1/(1-2\bar s)} \qquad
\rho \approx \frac{u_0 \bar s}{4\pi r_0^2} \left(\frac{r_0} {r}\right)^{2\bar s +1}.
\ee
 Therefore, the limit $|u|\to \infty$ describes the infinite/zero radius limit for type I/II.
For the type I, 
the radius goes to infinity and the density goes to zero, which is denoted as A3 in Table~\ref{table2}.
For the type II, the radius goes to zero as $u\to \infty$. 
When $-1/2<\bar s \leq 0$ i.e.
$ -(1+3w_1)/4 \leq w_2 < -(1+w_1)^2/4 -w_1/2$, the density at the origin diverges.
On the other hand, the density converges to zero or to a finite value when $\bar s \leq -1/2$ i.e.  $w_2 \geq -(1+w_1)^2/4 -w_1/2$. 
These behaviors are denoted as O7 and O8 in Table~\ref{table2}, respectively.

Let us consider a few specific cases.
For $w_2 = - (1+w_1)/4$, i.e. $s=0$ and $\bar s=(1-w_1)^{-1} $, Eq.~\eqref{g:v} can be reduced to 
\be{g:111}
g= \frac{1-u}{v} = \frac{1}{-\bar s+ |v/v_0|^{-(1-w_1)/(1+w_1)} }.
\ee
As $v\to \infty$, this gives asymptotic form $1-u = -(1-w_1)v$, where we use the explicit form for $\bar s$ in Eq.~\eqref{u:-inf}. 
The radius behaves as
$$
r  = r_0 |u|^{(1-w_1)/(1+w_1)} \to \infty,
$$
where $r_0$ is an integration constant which represents a scale. 
This case will be included in the limit C.
When $w_2 = -(1+w_1)^2/4$ i.e. $s=1/2=\bar s$,  Eq.~\eqref{g:v} is reduced to 
\be{w2=2}
|g|=\left|\frac{1-u}{v}\right| = \left(\frac{v}{v_0} \right)^{(1-w_1)/(1+w_1)}
 \quad \Rightarrow \quad v = v_0 \left| \frac{1-u}{v_0}\right|^{(1+w_1)/2}.
\ee
When $|u| \gg |v| \gg 1$, the radius behaves as $d\log r = du/(2v-u) \simeq -d\log u$. 
This gives $r  = |u_0/u| \ll 1$.
Therefore, this case presents the small $r$ region. 
This case will be included in the limit B.

\begin{figure}[bht]
\begin{center}
\begin{tabular}{c}
\includegraphics[width=.5\linewidth,origin=tl]{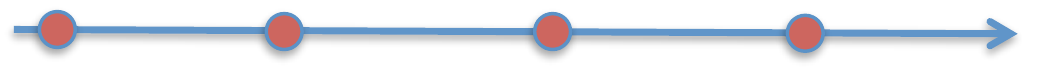} \\
\end{tabular}
\put (-275, 15) {$\bar s =$} \put(-275, 30) {$s =$} \put(-280,-10) {$w_2=$}
\put (-255, -10) {$\frac{3w_1-1}{4}$} \put (-255, 30) {$\frac{-2}{1+w_1}$}
\put (-205, -10) {$-\frac{1+w_1}{4}$} \put (-196, 15) {$ \frac{1}{1-w_1}$}  \put (-190, 30) {$0$}
\put (-148, 30) {$1/(1+w_1) $} \put (-126, 15) {$ 0$}
\put (-140, -10) {$-\frac{1+3w_1}{4}$}
\put (-92, -10) {$-\frac{(1+w_1)^2+2w_1}{4}$} \put (-72, 13) {$ -1/2$}
\put (-40, 30) {$ \to \infty$}  \put (-40, 15) {$ \to -\infty$}
\put (-8, 3) {$w_2$}
\put (-105, 0) {\tcr{\huge $\bullet$}} \put( -102, -10) {\tcr{0}}
\end{center}
\caption{Important values of $w_2$, $s$ and $\bar s$ for the classification of the asymptotic behaviors. 
The red dot represents the position of $w_2=0$ when $-1/3< w_1< 0$. 
}
\label{fig:sbar}
\end{figure}
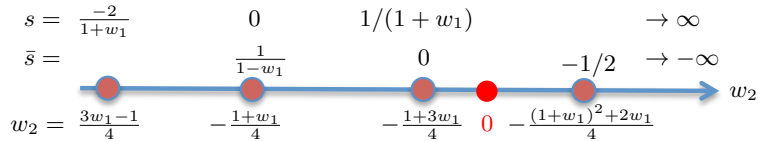

Various important values in the classification of the asymptotic properties are compared in Fig.~\ref{fig:sbar}.
For a given specific values of $w_1$ and $w_2$, the behaviors of a specific solution in the asymptotic limit can be consulted in table~\ref{table2}.   
\begin{table}[htp]
\caption{The behavior of a solution curve for large $|u|$}
\begin{center}
\begin{tabular}{|c|c|c|c|c|c|}
\hline 
	Role  & 
	Position &
	Condition & 
	$r/r_0$ & 
	$u-v$ & 
	$\rho$ 
	\\
\hline
 O3 &
	 $|u|\gg 1, ~v\sim 0$ &
	$ w_2 \leq \frac{3w_1-1}{4}$ & 
	$ |u|^{-1}$ & 
	$ v\propto |u|^{(1+w_1+4w_2)/(-2w_1)}  $ &
	$  r^{-\frac{1- 3w_1+4w_2}{-2w_1}} \to \rho_0  $ \\
\hline
 O4 &
	 $|u|\gg 1, ~v\sim 0$ &
	$ \frac{3w_1-1}{4}< w_2 < \frac{-1-w_1}{4}$ & 
	$ |u|^{-1}$ & 
	$ v\propto |u|^{(1+w_1+4w_2)/(-2w_1)}   $ &
	$  r^{-\frac{1- 3w_1+4w_2}{-2w_1}} \to \infty $ \\
\hline
   O5 &
	$|u|\gg 1$ &
	$ -\frac{1+w_1}{4}< w_2< -\frac{1+3w_1}{4}$ & 
	$ |u|^{-1}$ & 
	$ v\propto |u|^{(1+w_1+4w_2)/(-2w_1)}  $ &
	$  r^{-\frac{1- 3w_1+4w_2}{-2w_1}} \to \infty  $ \\
\hline
    O6 & 
 	 $|u|\gg1, ~v\sim 0$&
	$w_2> -\frac{1+3w_1}{4}$ & 
	$ |u|^{-1}$  & 
	$ v\propto |u|^{(1+w_1+4w_2)/(-2w_1)} $ & 
	$ r^{-\frac{1- 3w_1+4w_2}{-2w_1}} \to \infty  $\\
\hline
    O7 & 
	Asymptotic line & 
	$ -1/2 <\bar s \leq 0$ & 
	$(u_0/u)^{1/(1-2\bar s)} $ & 
	$g \to -\bar s^{-1} >0 $ & 
	$r^{-2\bar s-1}
		\to \infty $\\
\hline
    O8 & 
	Asymptotic line & 
	$ \bar s \leq -1/2$ & 
	$(u_0/u)^{1/(1-2\bar s)} $ & 
	$g \to -\bar s^{-1} >0 $ & 
	$r^{-2\bar s-1}
		\to \rho_0 $\\
\hline
  A3 &
  	Asymptotic line &
	$ \bar s > 1/2$ &
	$ (u_0/u)^{1/(1-2\bar s)}$ &
	$ g \to -\bar s< 0 $ &
	$r^{-2\bar s-1} \to 0 $ \\
\hline
\end{tabular}
\end{center}
\label{table2}
\end{table}%

\section{Exactly solvable case}
To understand the situation better, we present solution curves for the case with $w_1 = -1/3=w_2$, because this case allows exact solutions given in Ref.~\cite{Cho:2016kpf}.
This case belongs to type I and allows a linear solution curve in Eq.~\eqref{linear sol}. 
The values of $s$ and $\bar s$ are given by $-3/2$ and $3/2$, respectively.
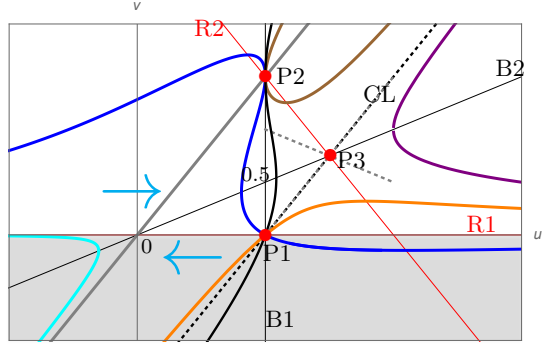
\begin{figure}[htb]
\begin{center}
\begin{tabular}{c}
\includegraphics[width=.4\linewidth,origin=tl]{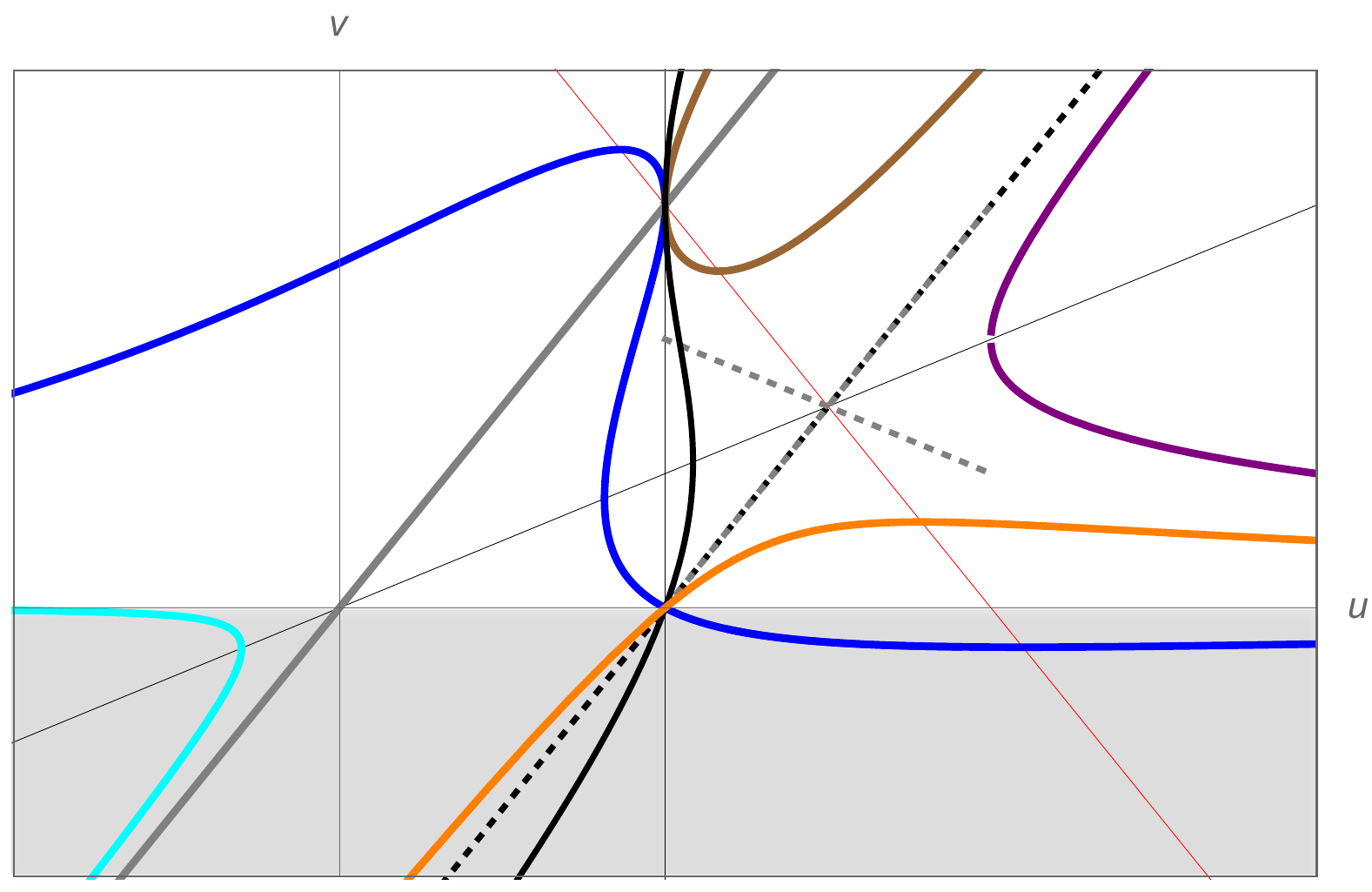}
\end{tabular}
\put (-154,-27) {\footnotesize $0$  }
\put (-119,0) {\footnotesize{ 0.5} }
\put (-20, 40) {B2}
\tcr{\put (-31, -19) {R1}
\put (-134, 55) {R2}}
\put (-107, -55) {B1}
\put (-106,37) { P2} \put (-110,37) {\tcr{\large $\bullet$}}
\put (-108,-30) {P1} \put (-110,-23.5) {\tcr{\large $\bullet$}}
\put (-80,5) {P3} \put (-85.5,7) {\tcr{\large $\bullet$}}
\put (-170, -10) {\textcolor{cyan}{ \Huge $\rightarrow $  }}
\put (-147, -35) {\textcolor{cyan}{ \Huge $\leftarrow $  }}
\put (-70, 30) {CL}
\end{center}
\caption{
Solution curves for type I with $w_1=-1/3= w_2$ i.e., $s= -3/2$.
Each curve of a given color represents a specific kinds of solution curve.
The characteristic lines R1, R2, B1, B2 are plotted as a dashed lines. 
The black dotted line denotes the characteristic line (CL) given in Eq.~\eqref{charline}.
The gray regions denotes unphysical region where the energy density is negative definite.
However, in the right upper corner with $u>1$ and $v>0$, $r$ plays the role of time and $-p_r= \rho/3$ is positive definite. 
The cyan arrows represent the increasing direction of $r$, which is divided by the line B2. 
The gray dotted lines around P3 represent the line $X_\pm$ in Eq.~\eqref{X:asym}. }
\label{fig:exact}
\end{figure}
The solution were shown to be divided into four different types. 
The S$_3$-I type solution is given in Eq.~\eqref{bh1}.
The solution contains two independent parameters $K$ and $R_0$. 
The corresponding solution curve is a blue one in Fig.~\ref{fig:exact}.
At the south pole, the geometry begins with an initial singularity at $u\to \infty$ (O3), where the abbreviation inside the bracket represents the behaviors of the solution curve around the region in Table~\ref{table1} and \ref{table2}. 
As $r$ increases, the geometry becomes static when the curve passes the point P1. 
Then, the radius takes its maximal value at P3.
The radius bounces back to decrease as $u\to -\infty$ where a time-like naked singularity appears at the north pole (O3).   
By changing the values of $K$ and $R_0$, indefinitely many different solution curves can be plotted which show qualitatively similar behavior to the blue one.
One of the two independent degrees of freedom in the parameters will be fixed by the choice of the solution curve.
The other freedom, which usually determines the scale of the solution, is determined when one integrates Eq.~\eqref{dxi}.

The S$_3$-II solutions are given by $\rho(\chi)= \frac{3}{8\pi R_0^2} (1\mp K \tanh \chi)$, 
where $r= R_0 \cosh\chi$.
The metric becomes 
$$
ds^2 =  (1 \mp K \tanh \chi) dt^2 +
	\frac{R_0^2}{-(1 \mp K \tanh \chi)} d\chi^2 + R_0^2 \cosh^2 \chi d\Omega_2^2 . 
$$
The $-$ solution corresponds to a blackhole-like solution given by the black curve in Fig.~\ref{fig:exact}. 
In the time-like region with $u> 1$ and $v> 0$, the solution curve begins with A3 asymptotic form around the line \As. 
$r$ decreases as the curve approaches P2.
At P2, the value of $r$ bounces back to increase until the curve goes to the limit $u\to -\infty$ and $v \to -\infty$ (A3). 
Meanwhile, the curve passes P1, which plays the role of a horizon. 
The $+$ solution corresponds to a cosmological solution, which is given by the brown curve.  
Each ends follow the \As limit (A3). 
The value of $r$ bounces at P2.  

The H$_3$ solution are given by $\rho(\chi)= -\frac{3}{8\pi R_0^2} (1\mp K \tanh \chi)$, 
where $r= R_0 \sinh\chi$.
The metric becomes 
$$
ds^2 = - (1 \mp K \coth \chi) dt^2 +
	\frac{R_0^2}{1 \mp K \coth \chi} d\chi^2 + R_0^2 \sinh^2 \chi d\Omega_2^2 . 
$$
The $-$ solution with $K<1$ corresponds to a black hole solution, which is given by an orange curve in Fig.~\ref{fig:exact}. 
The curve begins at $u\to \infty$ with $r=0$ (O3). 
As $r$ increases, it passes P1 and then follows the \As~line (A3). 
The $-$ solution with $K> 1$ corresponds to a non-static cosmological solution, which is given by the purple curve.
The curve also begins at $u\to \infty$ (O3) with $r=0$. 
The value of $r$ monotonically increases as the curve increases the \As~line (A3). 
The $+$ solution corresponds to a singular static solution, which is given by the cyan curve. 
The energy density for the solution is negative definite.
The curve begins at $u\to -\infty$ (O3) with $r=0$. 
As $r$ increases, the curve follows the \As~line (A3). 

A linear solution curve (the gray line) in Eq.~\eqref{linear sol} passes both the origin $\mathcal{O}$ and P2 linearly.
Because any solution curve never cross each other at points other than $\mathcal{O}$, P1, and P2, the gray line can be used to characterize the behavior of other solution curve.
For example, a blue-like curve may not pass the gray line at points other than P2.   
This implies that there is absent of a black hole solution in a closed spacetime which does not have a naked singularity.  
The brown curve will always be located at both sides of the gray line. 

\section{Numerical solutions}
In this section, we display various solution curves of the equation~\eqref{de2} on the $(u,v)$ plane after dividing the equation to types I, II, and III.
Characteristic forms of $C$ are displayed for each cases by choosing appropriate parameters for $(w_1,w_2)$.
Because we are mainly interested in black hole solutions in a closed space, main properties of which are determined at points P1, P2 and $\mathcal{O}$, the solution curves are classified based on the Table~\ref{table1}. 

The behaviors of a differentiable solution curve $C$ around P1 is described by Eq.~\eqref{SC:P1}. 
For $w_1=-1/3$, $C$ passes P1 linearly.
Therefore, $C$ in the region $(u>1,v> 0)$ will go into the region $(u<1,v<0)$ through P1. 
For $w_1 \gtrless -1/3$, $C$ passes P1 horizontally/vertically. 
Therefore, for $-1/3<w_1<0$, $C$ in the region $(u< 1,v>0)$ goes into the region $u> 1$ through P1.
However, the curve cannot go into the region $(u< 1,v< 0)$ through P1.  

The behavior of a solution curve around P2 should be interpreted in connection with the linear solution curve given in Eq.~\eqref{linear sol}, which connects P2 and $\mathcal{O}$.
In a general case other than $w_2 = -(1+w_1)/2$, the linear solution at $\mathcal{O}$ in Eq.~\eqref{origin} does not match with the linear solution at P2 in Eq.~\eqref{hor2}.
The linear solution~\eqref{origin} at $\mathcal{O}$ will be bent as $(u,v)$ departs from $\mathcal{O}$ and then will passes the point P2 vertically from above, or from the bottom, or may not pass P2 depending on the value of $w_2$.   
The regularity of the point $\mathcal{O}$, which plays the role of an origin $r=0$, is determined by the size of $w_2$ relative to $w_1$.
Therefore, $\mathcal{O}$ is singular/regular when $w_2\gtrless w_1$, respectively.

\subsection{Type I:  $w_2 \leq -(w_1+1)^2/4$ }
Let us consider the type I in Eq.~\eqref{cases}.
The values of $s$ and $\bar s$ satisfy $s\leq 1/2$ and $\bar s \geq 1/2$, respectively. 
As in Fig.~\ref{fig:pts}, the Type I system can be divided into three different classes, i) $w_2\leq w_1$; ii) $w_1 < w_2 < -(1+w_1)/2$; and iii) $-(1+w_1)/2 \leq w_2 \leq -(1+w_1)^2/4$.
The behaviors of a solution curve belonging to a specific class can be consulted in the table~\ref{table1}.
The system may be further classified by means of the asymptotic behaviors given in Table~\ref{table2}, which we do not pursue in this work.
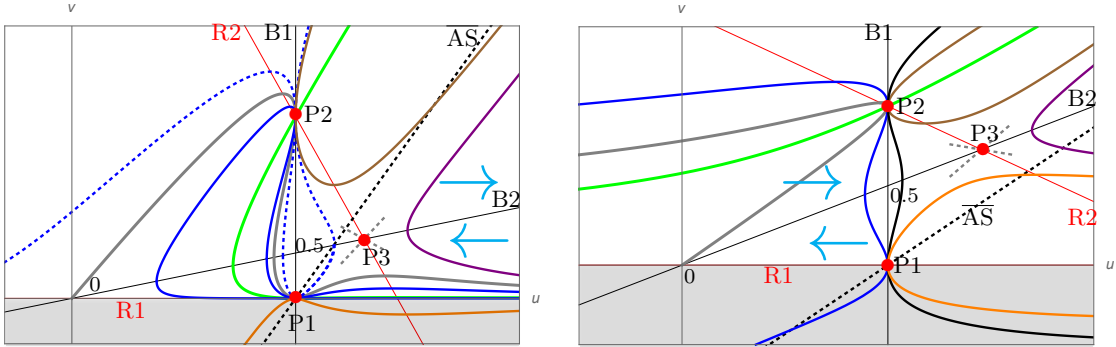
\begin{figure}[hbt]
\begin{center}
\begin{tabular}{cc}
\includegraphics[width=.4\linewidth,origin=tl]{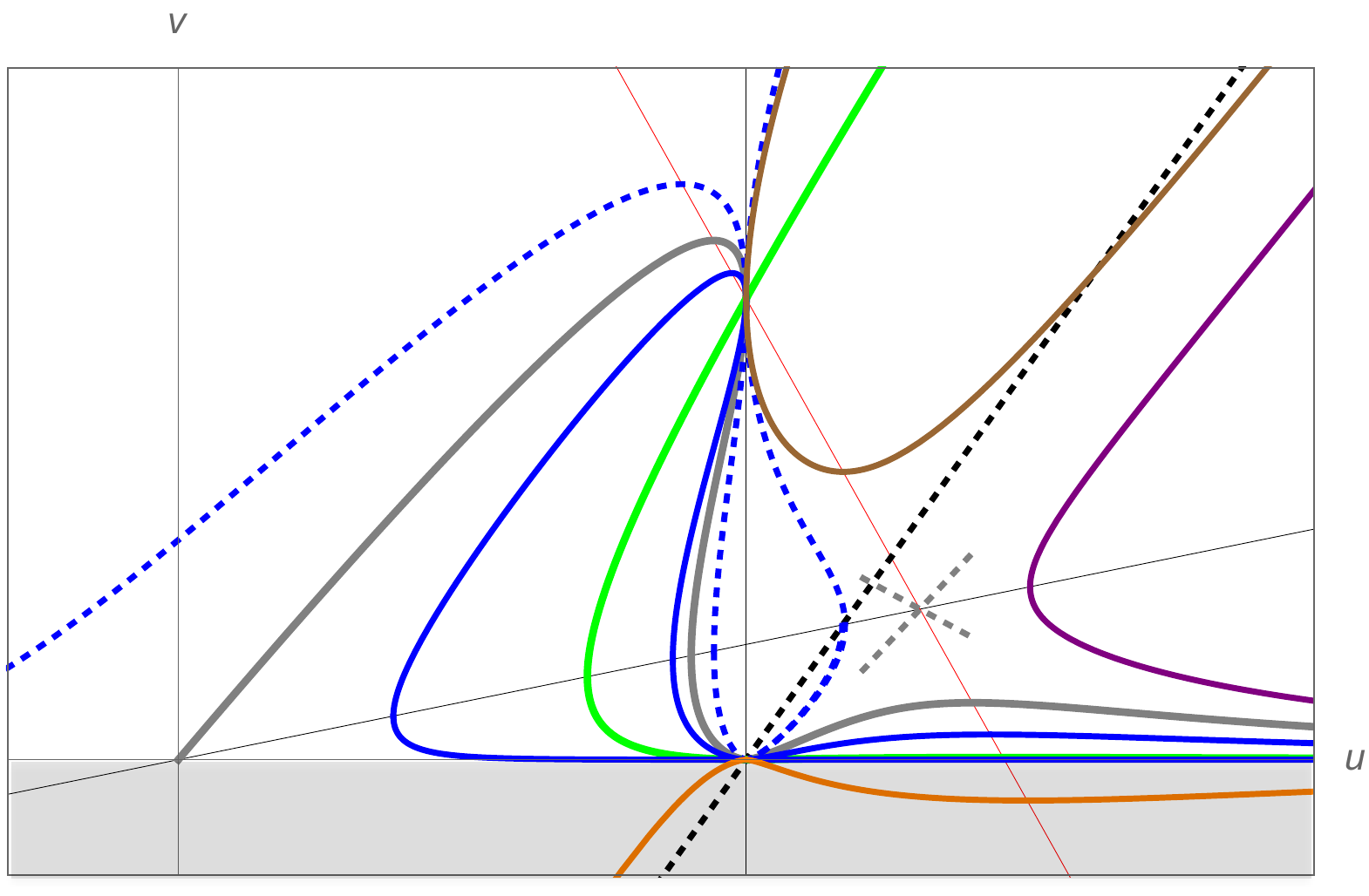}
&
\quad
\includegraphics[width=.4\linewidth,origin=tl]{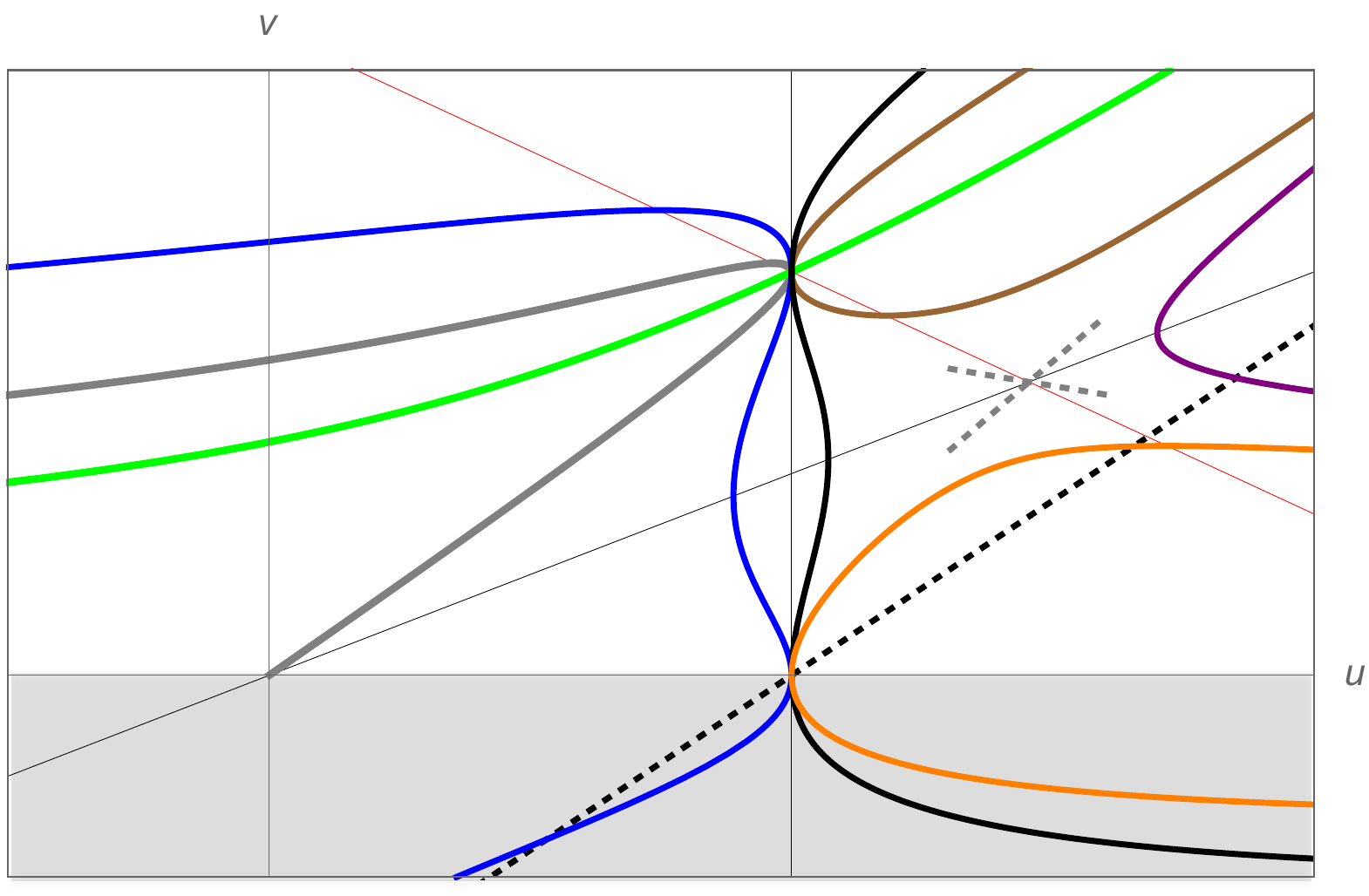}
\end{tabular}
\put (-390,-40) {\footnotesize $0$  }
\put (-315,-25) {\footnotesize{ 0.5} }
\put (-238, -8) {B2}
\tcr{\put (-380, -50) {R1}
\put (-344, 55) {R2}}
\put (-324, 55) {B1}
\put (-312,24) { P2} \put (-315,24) {\tcr{\large $\bullet$}}
\put (-315,-55) {P1} \put (-315,-45.3) {\tcr{\large $\bullet$}}
\put (-290,-30) { P3} \put (-289,-23.5) {\tcr{\large $\bullet$}}
\put (-259, -5) {\textcolor{cyan}{ \Huge $\rightarrow $  }}
\put (-255, -27) {\textcolor{cyan}{ \Huge $\leftarrow $  }}
\put (-255, 53) {\As}
\put (-165,-37) {\footnotesize $0$  }
\put (-90,-6) {\footnotesize{ 0.5} }
\put (-20, 30) {B2}
\tcr{\put (-135, -37) {R1}
\put (-20, -15) {R2}}
\put (-97, 55) {B1}
\put (-88,27) { P2} \put (-91,27) {\tcr{\large $\bullet$}}
\put (-89,-33) { P1} \put (-91,-33) {\tcr{\large $\bullet$}}
\put (-60,17) { P3} \put (-55,11) {\tcr{\large $\bullet$}}
\put (-129, -5) {\textcolor{cyan}{ \Huge $\rightarrow $  }}
\put (-119, -28) {\textcolor{cyan}{ \Huge $\leftarrow $  }}
\put (-60, -15) {\As}
\end{center}
\caption{
Typical forms of solution curves for type I. 
Here, $(w_1,w_2)=(-1/4, -1/2)$ i.e., $s= -10/3$ and $\bar s = 14/5$ (L) and  $(w_1,w_2)=(-1/2,-0.2)$ i.e., $s= -3/5$ and $\bar s = 13/15$ (R).
Each curve of a given color represents a specific kind of solution curve.
The characteristic lines R1, R2, B1, B2 are plotted as a dashed lines. 
The black dotted line denotes the asymptotic line (\As) given in Eq.~\eqref{charline}. 
The gray dotted lines around P3 represent the line $X_\pm$ in Eq.~\eqref{X:asym}. 
}
\label{fig:iso}
\end{figure}

When $w_2< 0$, the behavior of $C$ on the $(u,v)$ plane is qualitatively similar whether the density is singular or not at $\mathcal{O}$.
Because the case i) and ii) are distinguished by the behaviors of the density at $\mathcal{O}$ their solution curve will be similar. 
Therefore, we present two different sets of solution curves corresponding to 
(A) $(w_1>-1/3,w_2 < -(1+w_1)/2)$ and (B) $(w_1<-1/3,w_2\geq -(1+w_1)/2)$ for the classes i) and ii), respectively. 
As a specific example for each case, we choose $(w_1,w_2)=(-1/4, -1/2)$ and $=(-1/2, -0.2)$ for the cases (A) and (B), respectively. 
Examples of solution curve are given in Fig.~\ref{fig:iso}.
In each figure, except for the gray and the green curves, a given color curve is a representative of many different solution curves having similar characteristic.
On the other hand, the gray and the green curves are unique for a given $(w_1, w_2)$.
Along a solution curve $C$, from Eq.~\eqref{dxi}, the radius increases/decreases with $u$ at the upper/lower part than $\overline{\mbox{B2}}$ in the $(u,v)$ plane, which are denoted by the cyan arrows.

For the case (A), the {\it gray} solution curve which departs $\mathcal{O}$ linearly passes P2 vertically from the above as in the left panel of Fig.~\ref{fig:iso}. 
Then, the curve passes P1 with zero gradient. 
On the curve, the radius is maximized at P2 and the coordinate $r$ becomes timelike for $u> 1$. 
On the whole, this gray curve corresponds to a black hole solution, where a spacelike singularity at south pole is surrounded by a horizon at P1 and the north pole is regular. 
This solution curve complements the two deficits mentioned at the introduction just below of Eq.~\eqref{bh1}. 
There is no naked singularity and absent of negative energy density region.
The {\it green} curve corresponds to the linear solution~\eqref{hor2} around P2. 
As mentioned previously, the green curve appears as a limit of the quadratic solution~\eqref{hor}, the blue one.

The two (gray and green) curves can be used as a guide which characterizes the behaviors of other solution curves. 
For example, there is a {\it blue} solution curve which passes both points P1 and P2. 
Because the linear behavior at P2 is unique, any other solution curve which passes P2 should encompass the green curve around P2. 
Therefore, if the gray curve crosses P2 vertically from above as in this case, there exist solution curves which pass the horizon P1 twice.
One of such is plotted as the blue curve.  
Because the radius is maximized at P2, the blue curve represents a two black hole solution having two event horizons at both poles.  

The {\it dotted-blue} solution curve also has a horizon, i.e. passes P1.
Inside the horizon, a cosmological solution appears where the time ($r$ coordinate) will be minimized at P2.
Outside the horizon, a naked singularity appears at the north pole,  which behavior is similar to the blue curve in Fig.~\ref{fig:exact}.
There are other solution curves describing cosmological solutions.
The {\it brown} curve describes a cosmological solution bouncing at P2 and the {\it purple} curve describes an expanding universe which begins at $(u,v) \to (\infty,0)$ and ends around the asymptotic line given by the symbol \As.
The {\it orange} curve also describes a black hole. 
However, the solution is physically irrelevant because the energy density for the solution is negative definite. 

The class (B) solution curves are plotted in the right panel of Fig.~\ref{fig:iso}. 
The {\it gray} solution curve which departs $\mathcal{O}$ linearly passes P2 vertically from the bottom and then goes to $u \to -\infty$, which limit describes a naked singularity.
This solution describes a static closed spacetime where a star and a naked singularity are located at the south and the north poles, respectively. 
The {\it green} curve is located inside the gray one.
Now, every solution curve describing a static solution which passes P2 can be divided into two types: i) Solution curves inside the gray curve, of which curve is not shown in this figure. 
A corresponding solution will have naked singularities at both poles.
ii) Solution curves located outside of the gray curve, an example is the {\it blue} one. 
Every solution curve of this type passes P1, i.e., has a event horizon. 
Behind the point P1, the solution curve goes into negative energy region because $w_1<-1/3$. 
The other end of the blue curve goes to $u\to -\infty$, which describes a naked singularity because it should be located outside of the gray one.  
There are other cosmological solution curves like the {\it brown}, {\it purple}, and the {\it orange} curves. 
The behaviors of a solution curve in the asymptotic region follows the results in Table~\ref{table2}.

\subsection{Type II, ~~ $w_2\geq 0$}

As shown in the previous subsection, a solution curve passes P1 vertically when $-1<w_1< -1/3$.
This implies that a negative energy density region appears inevitably, which is unfavored due to the energy condition.
However, the geometry still can be used as a description of an outer part of a star.
To avoid the inevitable appearance of a negative energy density region, we restrict the value to $-1/3<w_1<0$ later in this work. 
In addition, we dismiss the stretched part of the solution curves to the negative energy regions. 

\begin{figure}[tb]
\begin{center}
\begin{tabular}{cc}
\includegraphics[width=.4\linewidth,origin=tl]{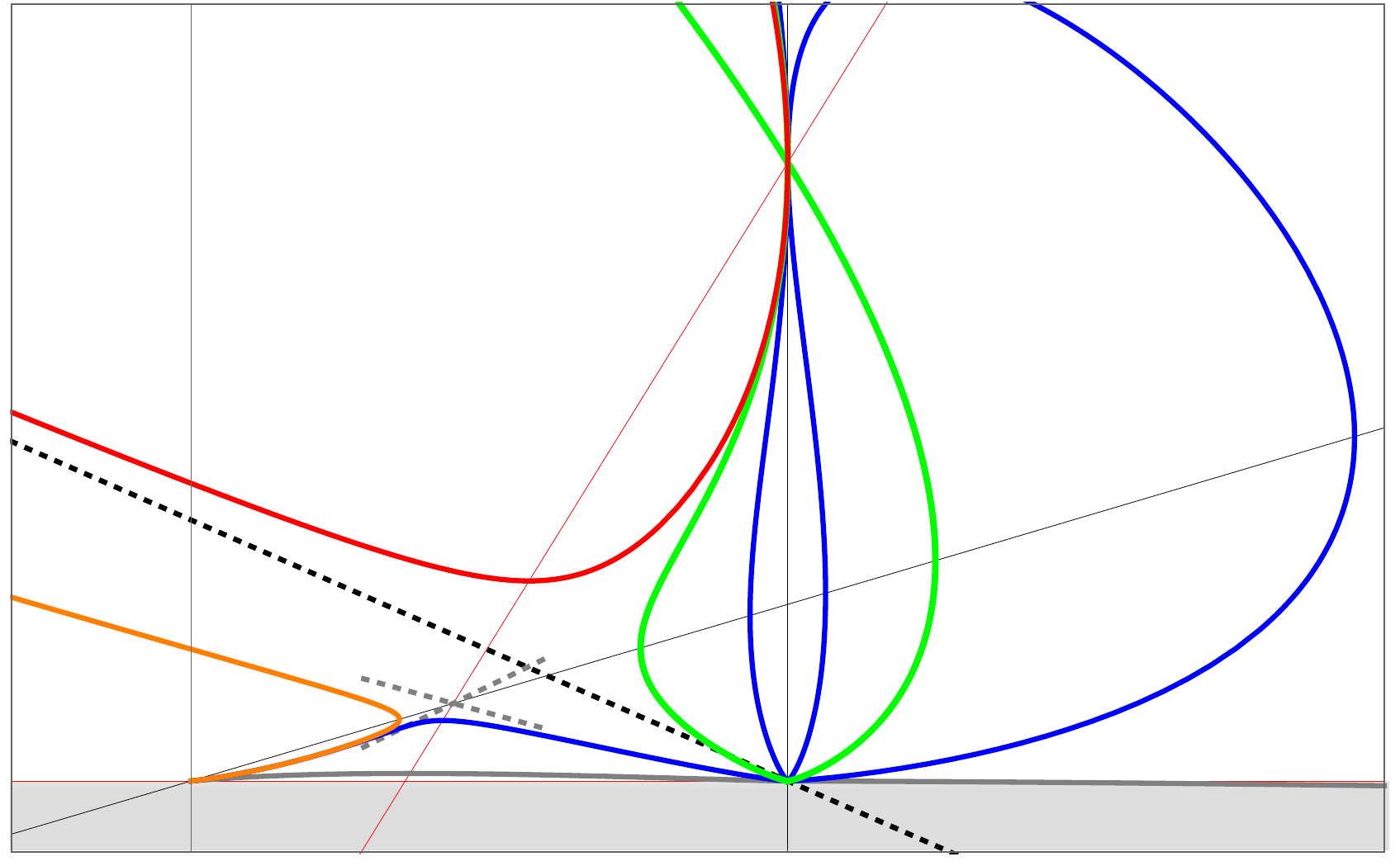}
&
\qquad
\includegraphics[width=.4\linewidth,origin=tl]{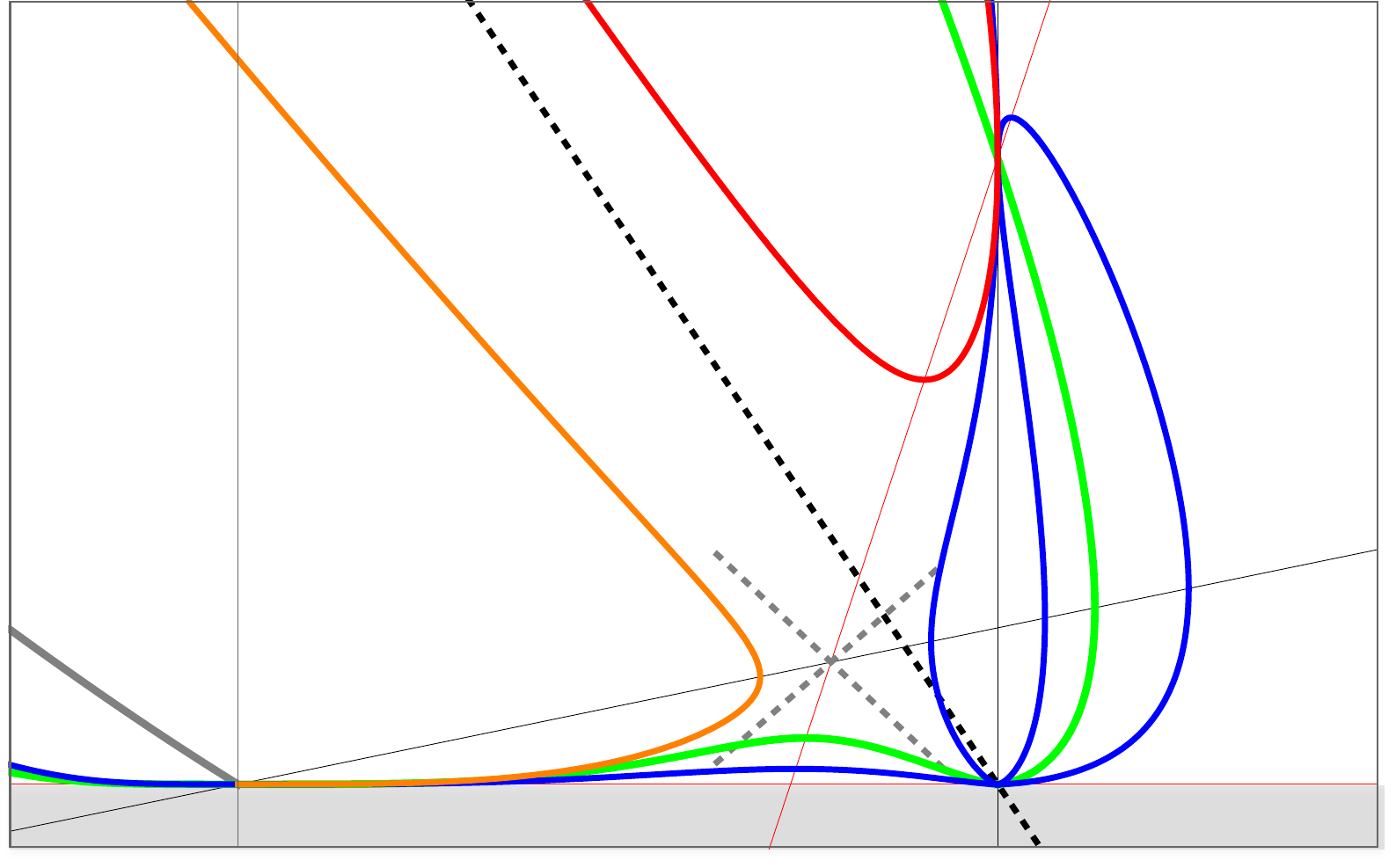}
\end{tabular}
\put (-409,-47) {\footnotesize $0$  }
\put (-325,-25) {\footnotesize{ 0.5} }
\put (-230, 2) {B2}
\tcr{\put (-245, -46) {R1}
\put (-310, 64) {R2}}
\put (-325, 64) {B1}
\put (-317, 40) { P2} \put (-320.5,40) {\tcr{\large $\bullet$}}
\put (-328,-54) {P1} \put (-320.5,-50) {\tcr{\large $\bullet$}}
\put (-376,-33) {P3} \put (-369.,-38.5) {\tcr{\large $\bullet$}}
\put (-280, -0) {\textcolor{cyan}{ \Huge $\rightarrow $  }}
\put (-280, -30) {\textcolor{cyan}{ \Huge $\leftarrow $  }}
\put (-420, -12) {\As}
\put (-170,-54) {\footnotesize $0$  }
\put (-66,-29) {\footnotesize{ 0.5} }
\put (-20, -13) {B2}
\tcr{\put (-25, -54) {R1}
\put (-54, 57) {R2}}
\put (-67, 64) {B1}
\put (-94,-26) {P3} \put (-87.5,-32.5) {\tcr{\large $\bullet$}}
\put (-76,41) { P2} \put (-63,41) {\tcr{\large $\bullet$}}
\put (-75,-54) { P1} \put (-63,-50.5) {\tcr{\large $\bullet$}}
\put (-33, 0) {\textcolor{cyan}{ \Huge $\rightarrow $  }}
\put (-30, -42) {\textcolor{cyan}{ \Huge $\leftarrow $  }}
\put (-127, 33) {\As}
\end{center}
\caption{Typical forms of solution curves for type II. 
Here, $(w_1,w_2) = (-2/7,1/10)$, i.e. $s= 0.273$ and $\bar s= -0.133$ (L) and $(w_1,w_2) = (-1/4,1/2)$, i.e. $s= 22/3$ and $\bar s = -18/5$ (R).
Each curve of a given color represents a specific kind of solution curve.
The characteristic lines R1, R2, B1, B2 are plotted as a dashed lines. 
The black dotted line denotes the asymptotic line (\As) given in Eq.~\eqref{charline}. 
In the left panel, the gray,  blue, and orange curves come from the limiting behavior in Eq.~\eqref{origin} with differences in the second order.   
}
\label{fig:caseII}
\end{figure}
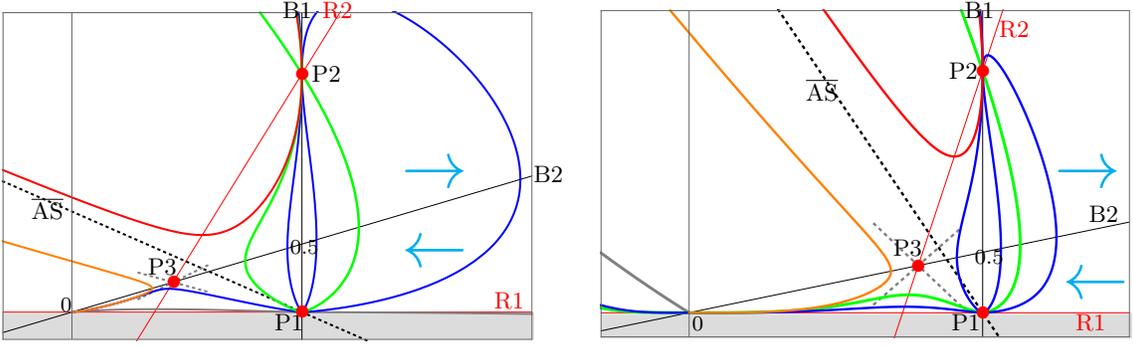

The values of $s$ and $\bar s$ satisfy $s \geq v_M$ and $\bar s \leq (1+ 3w_1)/(2w_1(1-w_1)) < 1/2$, respectively, for $w_2\geq 0$.
When $-1/3< w_1<0$, $\bar s$ is negative definite. 
Then a solution curve $C$ which follows the asymptotic line (\As) describes the near origin behavior $r \sim 0$.
The behavior of $C$ around $\mathcal{O}$ is divided into two types. 
For  $0<w_2  \leq -w_1/2$, the solution curve follows the linear behavior in Eq.~\eqref{origin}.
On the other hand, when $w_2> -w_1/2$, there are two different behaviors given in Eqs.~\eqref{origin} and \eqref{origin2}.
Therefore, we consider two different cases (A) $0<w_2  \leq -w_1/2$; (B) $w_2> -w_1/2$.
For both of the cases, the value of $w_1$ is restricted to $-1/3< w_1< 0$.  
Specifically for each case, we choose $(w_1, w_2) = (-2/7 ,1/10 )$ and $ = (-1/4, 1/2)$ for the cases (A) and (B), respectively.
Examples of solution curve are given in Fig.~\ref{fig:caseII}.
For the type II, the point $\mathcal{O}$ describes an asymptotic infinity $r\to \infty$.

Solution curves for the case (A) are plotted in the left panel of Fig.~\ref{fig:caseII}. 
Around $\mathcal{O}$, a solution curve has the linear behavior in Eq.~\eqref{origin} only.
However, this linear solution curve does not go to the point P2 but directly goes to the point P1, which curve is represented by the {\it gray} one between. 
However, a second order difference from the linear one (the gray one) allows that the corresponding solution curve behaves much differently from the gray curve at a distant point from $\mathcal{O}$.
Examples are the blue and the orange curves. 
%
Let us see the {\it green} curve, which is the linear solution at P2 in Eq.~\eqref{hor2}.
The radius $r$ takes a maximum value at P2 and decreases to zero as $u \to -\infty$. 
The green curve approaches the line \As~in Eq.~\eqref{charline} asymptotically.
From the other direction ($u>1$) of the green curve, the value of $r$ takes a minimum value at P2.
As the time coordinate $r$ increases, the curve approaches P1.
Then, it passes P1 to form an event horizon from the point of view of outside observers.
Subsequently, the radius $r$ takes a maximum value at P2 and starts to decrease as $u$ decreases.
Eventually, a naked singularity appears as the curve approaches the line \As ~(O7 or O8). 
A typical solution curve having (locally) maximum radii is the {\it blue} one. 
Even though it is not shown obviously in the figure, the origin ($r=0$) corresponds to the limit $(u,v) \to (-\infty, \infty)$, where a naked singularity appears when $-1/2 < \bar s < 0$.
The curve approaches the point P2 where the radius takes a (locally) maximum value.
Then, $r$ decreases and forms an event horizon where the curve passes P1.    
Behind the horizon, $r$ becomes a time coordinate and the metric describes a contracting anisotropic universe until it passes P2 once more. 
After that, the universe expands until it passes P1 once more to form another event horizon. 
Outside of P1, the radius increases until the curve arrives at the point $\mathcal{O}$, where \oo corresponds to the asymptotic infinity.
From the point of view of an outside observer residing in the asymptotic region, the segment \As-P2-P1 of the solution curve appears to describe a kind of baby universe behind horizons. 
Other than the blue one, there are two kinds of solution curves, the {\it orange} and the {\it red} ones.  
The orange curve begins around the asymptotic line with a naked singularity at $r=0$.
As $r$ increases, the curve approaches \oo asymptotically.
Both limits of the red curve correspond to naked singularities.  

Now, let us explain the right panel which corresponds to the case (B). 
A crucial difference from the left panel is that there exist two limiting behaviors at $\mathcal{O}$.
The linear one~\eqref{origin} is described by the {\it gray} curve which extends to $u\to -\infty$.
On the other hand, the power law one~\eqref{origin2} are extended to the positive $u$.
The {\it green} curve behaves similarly to that of the left panel until it passes the horizon. 
Outside the horizon, the curve approaches \oo and the metric describes an asymptotic region. 
The {\it blue}, {\it red}, and {\it orange} curves show similar behaviors with those in the left panel characteristically.

\subsection{Type III, ~~ $-(1+w_1)^2/4 \leq w_2 < 0$}

Finally, we display solution curves for the type III with $1/2 \leq s < v_M$ and $  (1+3w_1)/(2w_1(1-w_1)) < \bar s \leq 1/2$. 
We also restrict $w_1$ to $-1/3< w_1 < 0$ to protect from appearance of a negative energy density region. 

As discussed just below Eq.~\eqref{charline}, the asymptotic line does not play an important role in this case because solution curves do not converge on the line for large $|u|$ and $|v|$. 
The behaviors of a solution curve around $\mathcal{O}$ is given by Eqs.~\eqref{origin} where \oo plays the role of an origin $r=0$. 
Specifically, we choose $(w_1, w_2) = (-1/4 ,-1/30 )$.
Examples of solution curve are given in Fig.~\ref{fig:caseII}.

The linear solution curve at the point~\oo in Eq.~\eqref{origin} is plotted as the gray curve.
After the gray curve departs from the point, the curve passes P2 vertically from the bottom and then approaches $(u,v) \to (-\infty,\infty)$, where a naked singularity exists.
The linear solution curve which departs P2~\eqref{hor2} (the green curve) to the direction of increasing $u$ approaches P1 as $r$ increases. 
An horizon will be formed at P1.  
The curve passes P2 once more to form a maximal radius surface. 
Then, it goes to $(u,v) \to (-\infty,\infty)$, where a naked singularity exists.
A typical solution curve having maximum radii in this case is the {\it blue} one. 
The solution curve begins with a naked singularity with zero radius at the south pole [$(u,v) \to (-\infty,\infty)$] and has a maximal radius at P2. 
After the bounce of the radius, an event horizon exists at P1. 
Outside of P1, a time dependent cosmological region exists where the size of the universe may shrink or expand. 
The solution curve enters into the static region once more through P1  and forms another event horizon.
After that, there appears another locally maximum radius surface at P2.
Then, another naked singularity appears at the north pole. 

\begin{figure}[tb]
\begin{center}
\begin{tabular}{c}
\includegraphics[width=.4\linewidth,origin=tl]{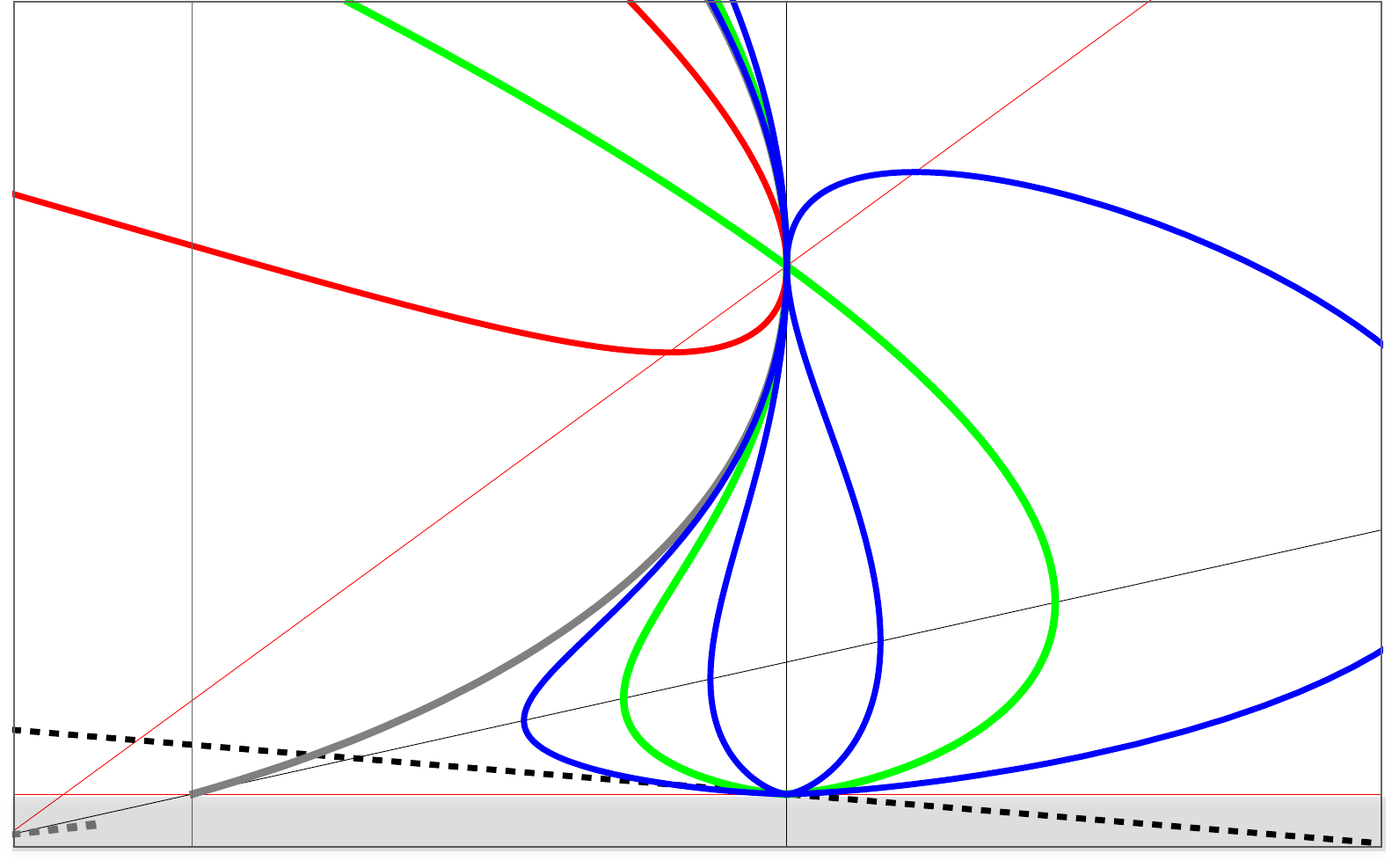}
\end{tabular}
\put (-177,-55) {\footnotesize $0$  }
\put (-100,-28) {\footnotesize{ 0.5} }
\put (-10, -17) {B2}
\tcr{\put (-145, -55) {R1}
\put (-48, 53) {R2}}
\put (-91, 58) {B1}
\put (-87,25.5) { P2} \put (-94,25.5) {\tcr{\large $\bullet$}}
\put (-93,-56) { P1} \put (-94,-52) {\tcr{\large $\bullet$}}
\put (-43, -10) {\textcolor{cyan}{ \Huge $\rightarrow $  }}
\put (-40, -32) {\textcolor{cyan}{ \Huge $\leftarrow $  }}
\put (-205,-37) {\As}
\end{center}
\caption{ Solution curves for the type III.
In this figure, we choose $w_1 = -1/4$ and $w_2 = - 1/30$, i.e. $s = 74/45$ and $\bar s= -14/75$.
}
\label{fig:caseIII}
\end{figure}
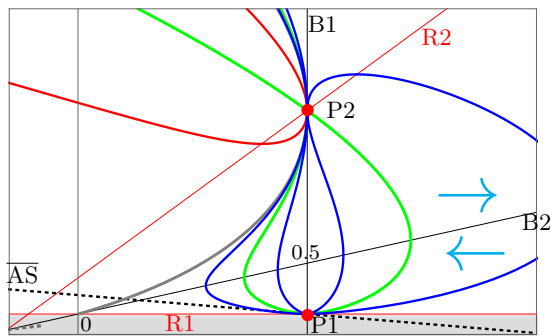

\section{Summary}
We have studied spherically symmetric geometries made of an anisotropic perfect fluid based on general relativity.
The angular pressure $p_2 = w_2 \rho$ may differ from the radial pressure $p_1 =w_1 \rho$, but because of spherical symmetry,
the pressure between different angular directions should be identical.   
To find and classify black hole solutions in closed space, 
we studied the conditions for a metric to form closed space and event horizon.  
In a general situation, we have found that a static surface of the locally maximal radius exists only when $-1 < w_1 < 0$.  

The Einstein equation for anisotropic fluids is eventually casted as a first-order autonomous equation in a two-dimensional plane of scale invariant variables $(u\equiv 2m(r)/r,v= 4\pi r^2 \rho)$.
The equation is equivalent to the TOV equation in general relativity.
We found that the autonomous equation is characterized by four specific lines, on which the integral curves of the equation are parallel to the axis $u$ or $v$.
Especially, the line $u=1$ (boundary of static/dynamical regions) and $v=0$ (boundary of the positive/negative energy regions) play important roles because the solution curves may not pass the lines without the use of specific points, i.e., $(u,v) = (1,0)$.  
The autonomous equation can be classified into three cases depending on the arrangement of the specific lines.
Then, we displayed various solution curves numerically. 

We first illustrated the behaviors of solution curves for the known exact solutions given in Ref.~\cite{Cho:2016kpf} for the case with $w_1 = -1/3=w_2$. 
One of the exact solutions describes a black hole solution in a closed spacetime. 
However, it bears two deficits, the appearances of a naked singularity and region of negative energy density.
These defects may not generally occur for an anisotropic perfect fluid, and we try to find explicit conditions to avoid them. 
We found that the negative energy density problem can be solved when $w_1$ is limited to $-1/3< w_1< 0$. 
In addition, there exists solution where the naked singularity can be hidden under a horizon when the fluid violates the strong energy condition.
Analytic solution corresponding to this case are not known.
However, the corresponding solution curve are plotted as a blue curve in the left panel of Fig.~\ref{fig:iso} and the physical properties in Table~\ref{table1} and~\ref{table2}.
In summary, we found that there exists black hole solutions without a naked singularity in closed space when the matter satisfies $\rho+3p_1>0$, and $\rho>0$ but violates the strong energy condition. 
For all other cases, at least one of the deficits survives.

Note that a regular star solution should begins at $\mathcal{O}(0,0)$ because its density must be finite and its mass should go to zero as $r\to 0$. 
The point $\mathcal{O}$ corresponds to the origin $r=0$ only when $w_2< 0$, which is satisfied with the case I and III.  
The corresponding solution curve is plotted as a gray curve in Figs.~\ref{fig:exact}, \ref{fig:iso}, and \ref{fig:caseIII}.
When $w_2< -(1+w_1)/2$, the curve extends to form a maximal radius surface, where the value of $v$ monotonically decreases and $u$ is held. 
Then, the radius bounces back to decrease.
At a smaller radius, an event horizon exists to hide a singularity behind.
 In this sense, it is depicted a combined system of a star and a black hole, which exist in the Arctic and Antarctic, respectively.
On the other hand, when $-(1+w_1)/2< w_2 < 0$, the curve extends to form a maximal radius surface where the value of $v$ monotonically increases and $u$ is held.  
At a smaller radius, the curve extends to the region with $u\to -\infty$.  
In this sense, the solution curve describes a combined system of a star and a naked singularity. 
In fact, there exists a regular star if $w_2 \leq w_1< 0$, where the solution curve begins at $\mathcal{O}$ with $r=0$.
For larger values of $w_2$, the point $\mathcal{O}$ becomes a singular origin or represents an asymptotic infinity. 
An interesting possibility of the present analysis is that there could exist static solutions between matters and black hole.
Therefore, the analysis may open a possibility to study a system having a black hole inside a star.

A quest to be done in the future is to improve the stability of the solutions, which was shown to be unstable in Ref.~\cite{Cho:2017nhx}.
The main origin of the instability is the negativity of the radial pressure.
There could be various way to avoid the instability such as introducing higher curvature gravity theory, other matters, and a cosmological constant.  
A most convenient way is to introduce a cosmological constant to an ordinary perfect fluid with a positive pressure, which gives negative equation of state naturally without introducing instability.

\section*{Acknowledgment}
This work was supported by the National Research Foundation of Korea grants funded by the Korea government NRF-2017R1A2B4008513 (HK).


\begin{thebibliography}{99}


\bibitem{Cho:2016kpf} 
  I.~Cho and H.~C.~Kim,
  Phys.\ Rev.\ D {\bf 95}, no. 8, 084052 (2017)
  doi:10.1103/PhysRevD.95.084052
  [arXiv:1610.04087 [gr-qc]].

\bibitem{Stephani2003}
Stephani H., Kramer D., MacCallum M., Hoenselaers C., Herlt E., {\it Exact Solutions of Einstein's Field Equations,} 2nd ed. Cambridge Monographs on Mathematical Physics, Cambridge University Press, New York (2003). 

\bibitem{Delgaty:1998uy} 
  M.~S.~R.~Delgaty and K.~Lake,
  Comput.\ Phys.\ Commun.\  {\bf 115}, 395 (1998)
  doi:10.1016/S0010-4655(98)00130-1
  [gr-qc/9809013].


\bibitem{Semiz:2008ny}
  I.~Semiz,
  Rev.\ Math.\ Phys.\  {\bf 23}, 865 (2011)
  doi:10.1142/S0129055X1100445X
  [arXiv:0810.0634 [gr-qc]].

\bibitem{Herrera1997}
Herrera L. and Santos N. O., Phys. Rep. {\bf 286}, 53 (1997).

\bibitem{Ruderman1972}
Ruderman R., Ann Rev. Astron. Astrophys. {\bf 10}, 427 (1972).

\bibitem{Bower1974}
R. L. Bower and E. P. T. Liang,
Astrophys. J. {\bf 188}, 657 (1974).



\bibitem{Matese1980}
J. J. Matese and P. G. Whitman,
 Phys. Rev. D {\bf 11}, 1270 (1980). 

\bibitem{Mak:2001eb} 
  M.~K.~Mak and T.~Harko,
  Proc.\ Roy.\ Soc.\ Lond.\ A {\bf 459}, 393 (2003)
  doi:10.1098/rspa.2002.1014
  [gr-qc/0110103].

\bibitem{Kippenhahn1990}
Kippenhahn R. and Weigert A., {\it Steller Structure and Evolution} Springer, Berlin (1990).

\bibitem{Dev2002}
Dev K. and Gleiser M., Gen. Re. Grav. {\bf 34} 1793 (2002).

\bibitem{Bhar2015}
Bhar P, Eur. Phys. J. C {\bf 75}, 123 (2015).   

\bibitem{Ratanpal:2016kwu} 
  B.~S.~Ratanpal and P.~Bhar,
  arXiv:1612.05417 [gr-qc].

\bibitem{Thirukkanesh2008}
Thirukkanesh S. and Maharaja S. D., Class. Quant. Grav. {\bf 25}, 235001-1 (2008).

\bibitem{Ivanov2002}
Ivanov B. V., Phys. Rev. D {\bf 65}, 104001-1 (2002).

\bibitem{Varela2010}
Varela V., Rahaman F., Ray S., Charkraborty K. and Kalem M., Phys. Rev. D {\bf 82}, 044052-1 (2010). 

\bibitem{Bekenstein:1971ej}
  J.~D.~Bekenstein,
  Phys.\ Rev.\ D {\bf 4}, 2185 (1971).
  doi:10.1103/PhysRevD.4.2185

\bibitem{Cho:2017nhx} 
  I.~Cho and H.~C.~Kim,
  arXiv:1703.01103 [gr-qc].


  

\end{thebibliography}
\end{document}